\documentclass[12pt,preprint]{aastex}

\received{21 December 2006}
\accepted{11 April 2007}
\journalid{\apj}
\usepackage{lscape}

\newcommand\iso[2]{$^{\rm #1}$#2}

\def\kmsec{\mbox{km~s$^{\rm -1}$}}
\def\teff{\mbox{T$_{\rm eff}$}}
\def\logg{\mbox{log $g$}}
\def\vt{\mbox{v$_{\rm t}$}}
\def\BmV0{\mbox{$(B-V)^{\rm 0}$}}
\def\VmK0{\mbox{$(V-K)^{\rm 0}$}}
\def\MV0{\mbox{$M_{\rm V}^{\rm 0}$}}

\def\etal{\mbox{et al.}}
\def\eg{\mbox{e.g.}}
\def\ie{\mbox{i.e.}}

\def\wnum{cm$^{-1}$}

\begin{document}

\title{
Improved Laboratory Transition Probabilities for Neutral Chromium and Re-determination of 
the Chromium Abundance for the Sun and Three Stars}

\author{
Jennifer S. Sobeck\altaffilmark{1},
James E. Lawler\altaffilmark{2}
and
Christopher Sneden\altaffilmark{1}
}

\altaffiltext{1}{Department of Astronomy and McDonald Observatory,
University of Texas, Austin, TX 78712; jsobeck@astro.as.utexas.edu,
chris@verdi.as.utexas.edu}

\altaffiltext{2}{Department of Physics, University of Wisconsin, Madison WI 53706; 
jelawler@wisc.edu}

%%%%%%%%%%%%%%%%%%%%%%%%%%%%%%%%%%%%%%%%%%%%%%%%%%%%%%
\begin{abstract}
%%%%%%%%%%%%%%%%%%%%%%%%%%%%%%%%%%%%%%%%%%%%%%%%%%%%%%

Branching fraction measurements from Fourier transform spectra in conjunction with 
published radiative lifetimes are used to determine transition probabilities for 263 lines of neutral chromium. 
These laboratory values are employed to derive a new photospheric abundance for the Sun: 
log $\epsilon$(Cr I)$_{\odot}$ = 5.64$\pm$0.01 ($\sigma = 0.07$).  These Cr I solar
abundances do not exhibit any trends with line strength nor with excitation energy and 
there were no obvious indications of departures from LTE.  In addition, oscillator strengths for singly-ionized 
chromium recently reported by the FERRUM Project are used to determine: 
log $\epsilon$(Cr II)$_{\odot}$ = 5.77$\pm$0.03 ($\sigma = 0.13$).  Transition probability data are also 
applied to the spectra of three stars: HD 75732 (metal-rich dwarf), HD 140283 (metal-poor subgiant), and CS 22892-052
(metal-poor giant).  In all of the selected stars, Cr I is found to be underabundant with respect to Cr II.
The possible causes for this abundance discrepancy and apparent ionization imbalance are discussed.
\end{abstract} 


%%%%%%%%%%%%%%%%%%%%%%%%%%%%%%%%%%%%%%%%%%%%%%%%%%%%%%
\section{INTRODUCTION}
%%%%%%%%%%%%%%%%%%%%%%%%%%%%%%%%%%%%%%%%%%%%%%%%%%%%%%

Accurate abundances are a key component necessary to understand stellar chemical evolution.
The derivation of reliable abundance values requires precise atomic data (i.e. radiative lifetimes,
branching ratios, and transition probabilities). Early laboratory efforts to determine these atomic parameters 
began decades ago.  Now with advances in astronomical
instrumentation, the accuracy of the atomic parameters has become a main source of uncertainty in the abundance determination. 
New investigations are underway which employ modern laboratory techniques 
to facilitate the extension and the improvement of these previous measurements of atomic data.  

Chromium is a member of the iron peak group $(Z = 24)$ with one dominant isotope (\iso{52}{Cr}).  
The synthesis of chromium is directly dependent on iron as the parent nucleus of \iso{52}{Cr} is
\iso{52}{Fe} (\eg\ Nakamura et al. 1999\nocite{Nak99}).  Prior to the mid-90's, abundance surveys
found [Cr/Fe] $\simeq$ 0 for stars across the entire range of metallicity 
(\eg\ Magain 1989\nocite{Mag89}, Ryan et al. 1991\nocite{Rya91}, Gratton \& Sneden 1991\nocite{Gra91}).
Taking into consideration both the nucleosynthetic linkage and the observational data, 
[Cr/Fe] was believed to remain at its solar ratio independent of metallicity.
However McWilliam \etal\nocite{McW95}\ (1995) examined abundances in a sample of extremely metal-poor stars,  
finding that [Cr/Fe] $\sim$ 0 until approximately [Fe/H] = $-2.5$, and then starts to 
decrease steadily with [Fe/H]. Additional observations by Cayrel et al. (2004\nocite{Cay04}) and Aoki et al. (2005\nocite{Aok05})
supported this finding.  Note that most of these abundance analyses employed only Cr I transitions.

The literature contains multiple studies of Cr oscillator strengths.  The major investigations include those of
Wujec \& Weniger (1981; Cr I/II),  Tozzi et al. (1985\nocite{Toz85}; Cr I) and Blackwell et al. (1984a\nocite{Bla84a}, 1986\nocite{Bla86}; Cr I).  
A new study of transition probabilities for singly-ionized chromium from the FERRUM Project 
has recently been published (Nilsson et al. 2006\nocite{Nil06}).  These Cr II results provided motivation for us to undertake a new
study of Cr I oscillator strengths.  Together, these two sets of high-quality transition probability data would allow for the 
accurate re-determination of the Cr abundance in the solar photosphere and examination of its ionization equilibrium.

In \S2, we briefly summarize the radiative lifetimes employed in our work.  
\S3 contains a description of our use of National Solar Observatory (NSO) digital archive spectra to 
measure Cr I branching fractions and a list of our new Cr I oscillator strengths. 
A report of our determinations of the Cr abundance in Sun and three other stars is found in \S4, along with a 
discussion of the implications of our Cr abundance analysis.  

%%%%%%%%%%%%%%%%%%%%%%%%%%%%%%%%%%%%%%%%%%%%%%%%%%%%%%
\section{Radiative Lifetime Measurement Summary}
%%%%%%%%%%%%%%%%%%%%%%%%%%%%%%%%%%%%%%%%%%%%%%%%%%%%%%

Cooper \etal\ (1997\nocite{Coo97}) reported radiative lifetime measurements for 131 levels of Cr I.  
They employed a time-resolved Laser Induced Fluorescence 
(LIF) technique on a slow atomic beam of Cr atoms from a 
hollow cathode discharge.   Cooper \etal\ were attentive to possible systematic errors from: (1) electronic bandwidth, linearity, and 
fidelity limitations, (2) flight-out-of view effects, (3) radiation trapping, (4) collisional quenching, and (5) 
Zeeman quantum beats.  Most importantly they re-measured certain "benchmark" lifetimes to check the accuracy of their 
apparatus during their work on Cr I.  Table 1 is a list of radiative lifetimes from Cooper \etal\ for the 65 Cr I levels 
included in our branching fraction study.

The $\pm$5\%\ accuracy claim of Cooper et al. (1997\nocite{Coo97}) may be verified by matching up their results to other (less
extensive) Cr I LIF measurements.  Specific comparisons of the average and the root mean square (RMS) differences between 
their lifetimes and other literature values are as follows: $+$0.9\%\ and 2.5\%\ respectively 
for three levels in common with the study by Measures et al. (1977\nocite{Mea77}), $+$8.5\%\ and 9.0\%\ respectively for six levels shared with 
the determination by Marek (1975\nocite{Mar75}), and $-$2.0\%\ and 5.1\%\ respectively for twenty-three levels in common with the study of 
Kwiatkowski et al. (1981\nocite{Kwi81}) \footnote{The reference values for these differences are the Cooper \etal\ data.}.  
The slightly larger discrepancy between the Cooper \etal\ lifetimes and those of Marek 
is not a concern as Marek claimed 8\%\ uncertainty on his measurements.  Two 
separate individual LIF lifetime measurements by Cooper et al. are in good agreement with values reported by Kwong \& 
Measures (1980\nocite{Kwo80}) and Hannaford \& Lowe (1981\nocite{Han81}).  In addition, their results compare relatively well to measurements 
with non-LIF techniques.  For example, six lifetimes determined by Marek and Richter (1973\nocite{Mar73}; phase-shift method) agree with 
the Cooper \etal\ measurements, as do six lifetimes measured by Becker et al. (1977\nocite{Bec77}; level-crossing technique).  

The comparison to the National Institute of Standards and Technology (NIST) critical compilation (Martin \etal\ 1988\nocite{Mar88}) 
in Table 1 involved the summation over all of the Einstein A-coefficients for the 
transitions from the upper level.  In some cases the sum is incomplete and only an upper limit can be determined.  
The NIST critical compilation included results from a variety of techniques. Although it is not expected to be as 
accurate as individual LIF measurements, the NIST compilation actually agrees well with the Cooper et al. results. 

%%%%%%%%%%%%%%%%%%%%%%%%%%%%%%%%%%%%%%%%%%%%%%%%%%%%%%%%%%%%%%%%%
\section{Branching Fractions and Atomic Transition Probabilities}
%%%%%%%%%%%%%%%%%%%%%%%%%%%%%%%%%%%%%%%%%%%%%%%%%%%%%%%%%%%%%%%%%

It was our intention to record new spectra on a variety of Fe-group species during a Kitt Peak run in June 2005.  
Maintenance problems with the NSO 1.0 m Fourier Transform Spectrometer (FTS) prevented the acquisition of new data 
and led to an extended shutdown of this crucial FTS facility. Despite this setback, we were able to move forward and generate
an expanded set of Cr I transition probabilities with existing FTS spectra.

The primary specifications of the FTS instrument are:
(1) wavenumber accuracy of 1 part in $10^{8}$, (2) broad spectral coverage range from the UV to the IR, (3) optimal resolution
limit of 0.01 \wnum, and (4) spectrum recording capability of $10^{6}$ points in 10 minutes (Brault 1976\nocite{Bra76}).  
The FTS interferogram is a simultaneous measurement on spectral resolution elements from the UV to near IR.  
This gives the FTS an advantage over single-channel, sequentially-scanned grating monochromators which 
are more vulnerable to branching fraction errors from drifts in source performance. 

We performed a search of all spectra in the publicly-accessible digital archives of the NSO.  Numerous chromium spectra were 
located, however based on various selection criteria (\ie\ range of lamp currents and relatively low
buffer gas pressures) only seven were deemed acceptable.   General characteristics of the 
spectral data set include: hollow cathode discharge (HCD) lamp sources
with fused silica windows, argon or neon fills, interferograms with multiple co-adds, applied currents in the range
0.1-1.5 A, and buffer gas pressures in the range of 0.65-4.00 Torr.  Details regarding the chosen spectra may be 
found in Table 2.  The majority of spectra were recorded at high discharge currents.  Though high current spectra
yield good S/N measurements of weak lines, radiation trapping or optical depth effects arise.  For instance, spectra 3-5 have 
serious problems of this sort even for emission lines terminating on Cr levels with 1 eV excitation potentials.  Spectrum 6, 
although not the lowest current spectrum, has minimal optical depth problems for Cr I lines to the 1 eV lower levels.  
In order to resolve these optical depth problems and to improve the radiometric calibration in the near IR, 
we made some additional laboratory measurements with a grating spectrometer which will be described in \S3.1.

Essential to the branching ratio measurement is the accurate determination
of the relative radiometric calibration or efficiency of the FTS.  In effect, the radiometric efficiency 
is the quantification of the FTS instrument response. A methodology to arrive at a 
radiometric calibration has been established 
by Adams \& Whaling (1981\nocite{Ada81}) which involves the use of selected sets of Ar I and Ar II lines in the 
range 4300-35,000 \wnum.  Confirmation and refinement of these lists of Ar I and Ar II branching
ratios have subsequently been done by Danzmann \& Kock (1982\nocite{Dan82}), Hashiguchi \&
Hasikuni (1985\nocite{Has85}) and Whaling et al. (1993\nocite{Wha93}).  The apparent intensity of subsets of
Ar I and Ar II lines from a common upper level divided by the branching ratios of these lines is used
to determine the FTS radiometric efficiency as a function of wavenumber.  The radiometric calibration 
includes efficiency variations as a function of wavenumber from the optical components of the 
FTS and lamp system.  Calibrations based solely on the Ar line technique were used for spectra 1, 2, and 6.  
Spectra of the Kitt Peak Optronics 15 A tungsten strip lamp were recorded during the 1984 run shortly before and 
after the Cr-Ne hollow cathode lamp spectra.  This tungsten filament lamp is a secondary standard with a known spectral 
radiance and its spectra were used to establish a relative radiometric calibration of Cr-Ne spectra 3, 4, and 5 from 1984. 
A spectrum of a 6.25 A tungsten filament lamp from March 25, 1980 was used to smooth the Ar line calibration of spectrum 
7 from the same date.  We do not have access to the calibration curve for this standard lamp, 
but we were able to reconstruct a calibration using other archived FTS data from the same period.
Note that tungsten filament lamp calibrations are most useful near the decline in FTS sensitivity at 
12,500 \wnum\ (attributed to the aluminum mirror coatings), 
and between 10,000 and 8,000 \wnum, where the response of the silicon detector rapidly diminishes.

All possible transition wavenumbers between the known energy levels of Cr I satisfying both the parity change and 
$\Delta J = 0, \pm 1$ selection rules were computed and used during the analysis of the FTS data.  Energy Levels 
from Sugar \& Corliss (1985\nocite{Sug85}) were used to determine all possible transition wavenumbers.  Spectral features
at these wavenumbers were numerically integrated to determine apparent line intensities that are subsequently
divided by the relative radiometric calibration to yield branching ratios.

The procedure for determining branching fraction uncertainties has been extensively described in Wickliffe et al. 
(2000\nocite{Wic00}). Branching fractions from a given upper level are defined to sum to unity, thus a dominant line from an 
upper level has small branching fraction uncertainty almost by definition. Branching fractions for weaker lines near the 
dominant line(s) tend to have uncertainties limited by their signal-to-noise ratios. Systematic uncertainties 
in the radiometric calibration are typically the most serious source of uncertainty for widely-spaced lines from a common upper level.

Branching fraction measurements were completed on 65 of the 131 levels from the lifetime experiment. Some of the levels for 
which branching fractions could not be obtained have significant branches which fall outside the spectral coverage region of the 
FTS configuration.  The division of the branching fractions by the radiative lifetimes results in the transition
probabilities for chromium lines.  Table 3 presents oscillator strengths for 263 transitions of Cr I.  
Note that the table omits transition probability data for a few weak lines from selected upper levels. 
These omissions are due to excessively large uncertainties from low S/N, blending issues, and some calibration 
problems for weak lines widely separated from dominant branches connected to the same upper level.  
Branching fractions of strong lines were corrected using our rough measurements on the omitted weak lines.  
Inaccuracies in the branching fractions of the weak lines have negligible effect on the accuracy of the branching 
fractions for the strong lines.  The branching fraction uncertainty was combined in
quadrature with the radiative lifetime uncertainty to yield the transition probability uncertainty shown in Table 3. 

%%%%%%%%%%%%%%%%%%%%%%%%%%%%%%%%%%%%%%%%%%%%%%%%%%%%%%
\subsection{Grating Spectrometer Measurements}
%%%%%%%%%%%%%%%%%%%%%%%%%%%%%%%%%%%%%%%%%%%%%%%%%%%%%%
        
Supplemental measurements were made using a 0.5m focal length grating spectrometer equipped with a set of dye 
and interference filters and a diode detector array.  The purpose of these supplemental measurements was to verify and improve
the IR radiometric calibration of the primary FTS data as well as further investigate the optical depth problems
in a small portion of this data.  We used small, sealed Cr-Ne and Cr-Ar hollow cathode lamps, which are 
standardly found in atomic absorption spectrophotometers.  To eliminate optical depth concerns during these measurements, the 
operational discharge current was limited to a range of 1.0-4.0 mA.  Two diffraction gratings were used: 
a first-order 1200 groove/mm grating for broad coverage in a single exposure 
and an echelle grating with 316 groove/mm and a 63 degree blaze for high spectral resolution.  
A tungsten-quartz-halogen lamp was employed to calibrate the radiometric response of the spectrometer system
(which included filters used to suppress scattered radiation).  Special attention was devoted to optical depth 
effects in the z$^{5}$P to a$^{5}$S and z$^{5}$P to a$^{5}$D multiplets.  In the solar Cr abundance determination 
of Blackwell \etal\ (1987\nocite{Bla87}), these multiplets were specifically mentioned and will be discussed in further detail in \S4.4. 

Spot checks with the grating spectrometer were performed on the the longest wavelength z$^{5}$F to a$^{5}$G multiplets.
These re-measurements indicated that in the IR, FTS data calibrated with
the Optronic 15 A tungsten strip lamp were more accurate than those calibrated with the Ar I/II line method. 
This is in part due to the weakness of the Ar I lines at high discharge currents.  However, the
overwhelming majority of final transition probabilities listed in Table 3 are derived from FTS spectra.

%%%%%%%%%%%%%%%%%%%%%%%%%%%%%%%%%%%%%%%%%%%%%%%%%%%%%%
\subsection{Theoretical Transition Probabilities}
%%%%%%%%%%%%%%%%%%%%%%%%%%%%%%%%%%%%%%%%%%%%%%%%%%%%%%

Some of the high spin levels of transition metals such as Cr display relatively pure LS or Russell-Saunders coupling.  
We have used the standard LS formulae from Condon \& Shortley (1935\nocite{Con35}) with a frequency cubed correction to compute transition 
probabilities for selected multiplets in Table 3.  The absolute scale of the transition probabilities of each multiplet 
was normalized to match one line in the multiplet.  The lines used for these normalizations are labeled with a "N" in Table 3. 
As shown, the A-values from LS-coupling computations agree well with those from experiment (where the average
and rms values of ($A_{Exp} - A_{LS calc})/A_{Exp}$ are found to be 0.011 and 0.084 respectively).

%%%%%%%%%%%%%%%%%%%%%%%%%%%%%%%%%%%%%%%%%%%%%%%%%%%%%%
\subsection{Comparison to Previous Studies}
%%%%%%%%%%%%%%%%%%%%%%%%%%%%%%%%%%%%%%%%%%%%%%%%%%%%%%

Numerous investigations of oscillator strengths for neutral chromium have been made with a variety of techniques. 
Significant experimental initiatives in this vein include the pioneering work of Corliss \& Bozmann (1962\nocite{Cor62}), the
shock tube approach of Wolnik et al. (1968\nocite{Wol68}, 1969\nocite{Wol69}), and the arc emission method of Wujec \& Weniger (1981\nocite{Wuj81}). 
Theoretical determinations of Cr I transition probabilities have been done by Bi\'{e}mont (1974\nocite{Bie74})
and Kurucz \&  Peytremann (1975\nocite{Kur75}).  Here, we focus on comparisons with two more recent sets of measurements: the
hook and emission method of Tozzi et al. (1985\nocite{Toz85}) and the furnace absorption technique of Blackwell et al. (1984a\nocite{Bla84a}, 
1986\nocite{Bla86}).  We also comment on the relationship of our gf data to those of the NIST critical compilation. 

\begin{figure}
\epsscale{0.80}
\plotone{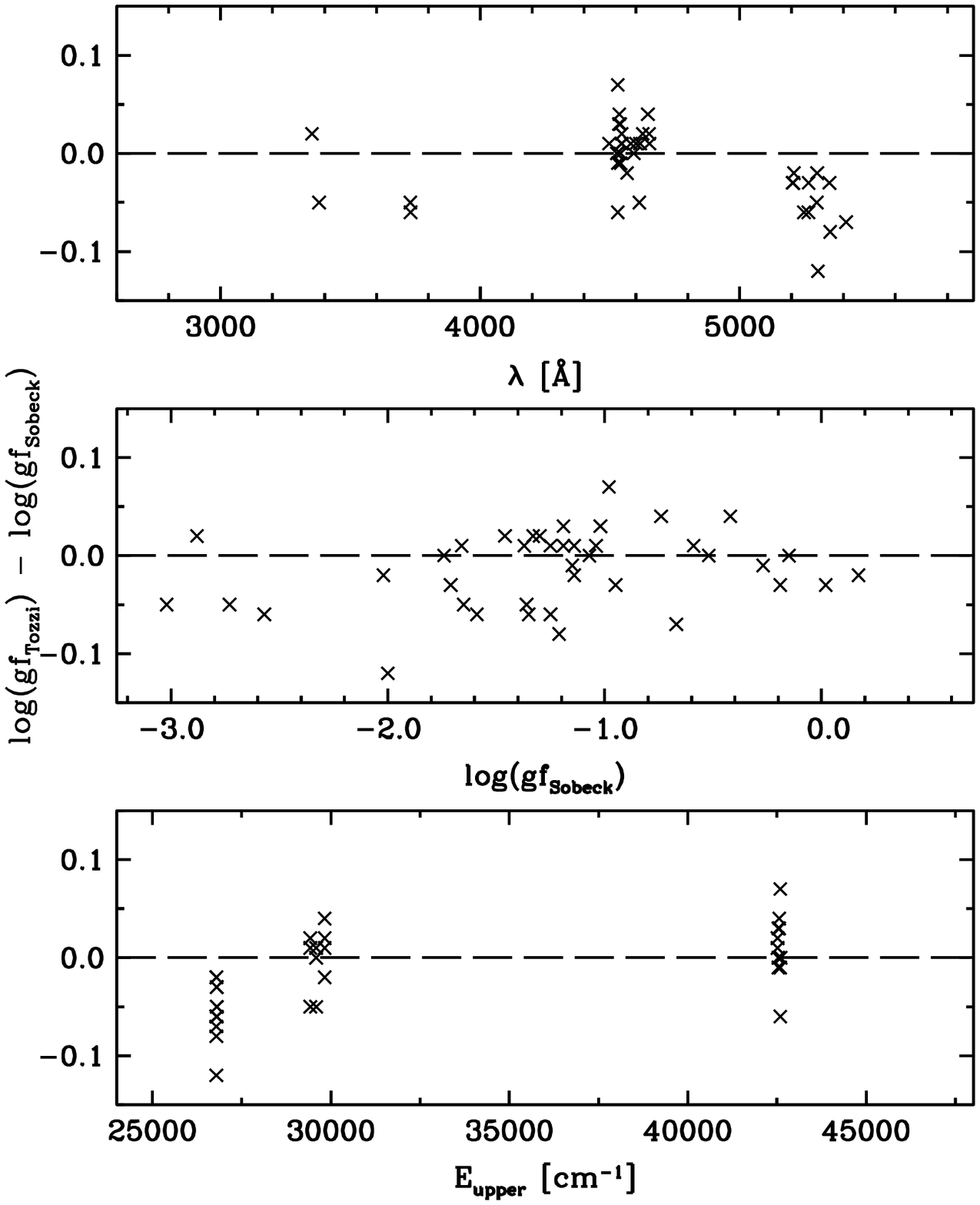}
\vspace*{0.2in}
\caption{Comparison of our oscillator strength values to those of Tozzi et al. (1985).  The upper panel shows the difference
         between the log($gf_{Tozzi}$) and log($gf_{Sobeck}$) as a function of wavelength.  
         The middle panel displays the difference verus the log($gf_{Sobeck}$) values.
         The bottom panel illustrates the difference as a function of upper energy level (E$_{upper}$). 
\label{f1}}
\end{figure}

Oscillator strengths for 60 Cr I lines (which originate from 14 different upper levels) were reported by Tozzi
\etal\ (1985\nocite{Toz85}).  Measurements of branching fractions were done with the hook and emission method.  
The Tozzi \etal\ branching fractions were combined with the radiative lifetimes from Kwiatkowski et al. (1981) 
to yield the gf values with an internal accuracy of 7\%.  
Figure~\ref{f1} shows the differences between the log gf values from Tozzi \etal\ (1985\nocite{Toz85}) and those from our work, as a function of 
wavelength, transition probability, and upper level transition energy.  Overall, the agreement between their values 
and ours is quite good.  A minor trend in wavelength is seen in the upper panel of Figure~\ref{f1}: the agreement between the two 
data sets worsens as wavelength increases.  The lower panel of Figure~\ref{f1} displays a similar trend with E$_{upper}$.  
We have 41 transitions in common with Tozzi \etal\ with the average 
and rms values of $log(gf)_{Tozzi} - log(gf)_{Sobeck}$ calculated to be $-$0.01 and 0.04 respectively.  

\begin{figure}
\epsscale{0.80}
\plotone{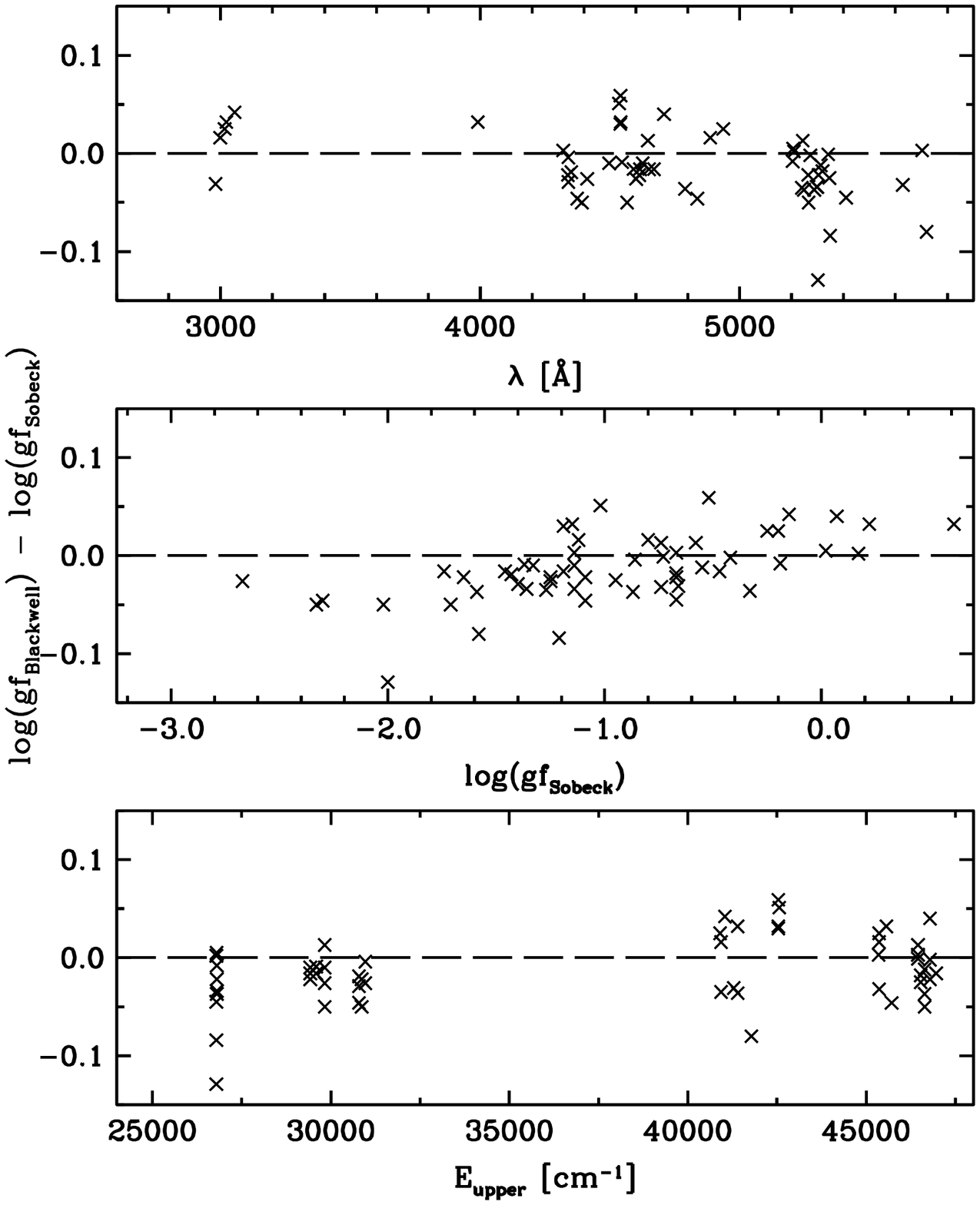}
\vspace*{0.2in}
\caption{Comparison of our oscillator strength values to those of Blackwell et al. (1984, 1986).  The upper panel shows the difference
         between the log($gf_{Blackwell}$) and log($gf_{Sobeck}$) as a function of wavelength.  
         The middle panel displays the difference verus the log($gf_{Sobeck}$) values.
         The bottom panel illustrates the difference as a function of upper energy level (E$_{upper}$). 
\label{f2}}
\end{figure}

With the use of the furnace absorption technique, the Blackwell group published two papers (1984a\nocite{Bla84a}, 1986\nocite{Bla86}) 
on Cr I oscillator strengths.  The yield of the two efforts was 102 Cr I oscillator strengths from 
38 different upper levels with an internal accuracy claim of better 
than 1\%.  Figure~\ref{f2} displays a comparison of our gf data to that of Blackwell et al.  There is excellent agreement between the 
57 lines that we have in common.  The average and rms values of $log(gf)_{Blackwell} - log(gf)_{Sobeck}$ 
are found to be $-$0.01 and 0.04 respectively.

\begin{figure}
\epsscale{0.80}
\plotone{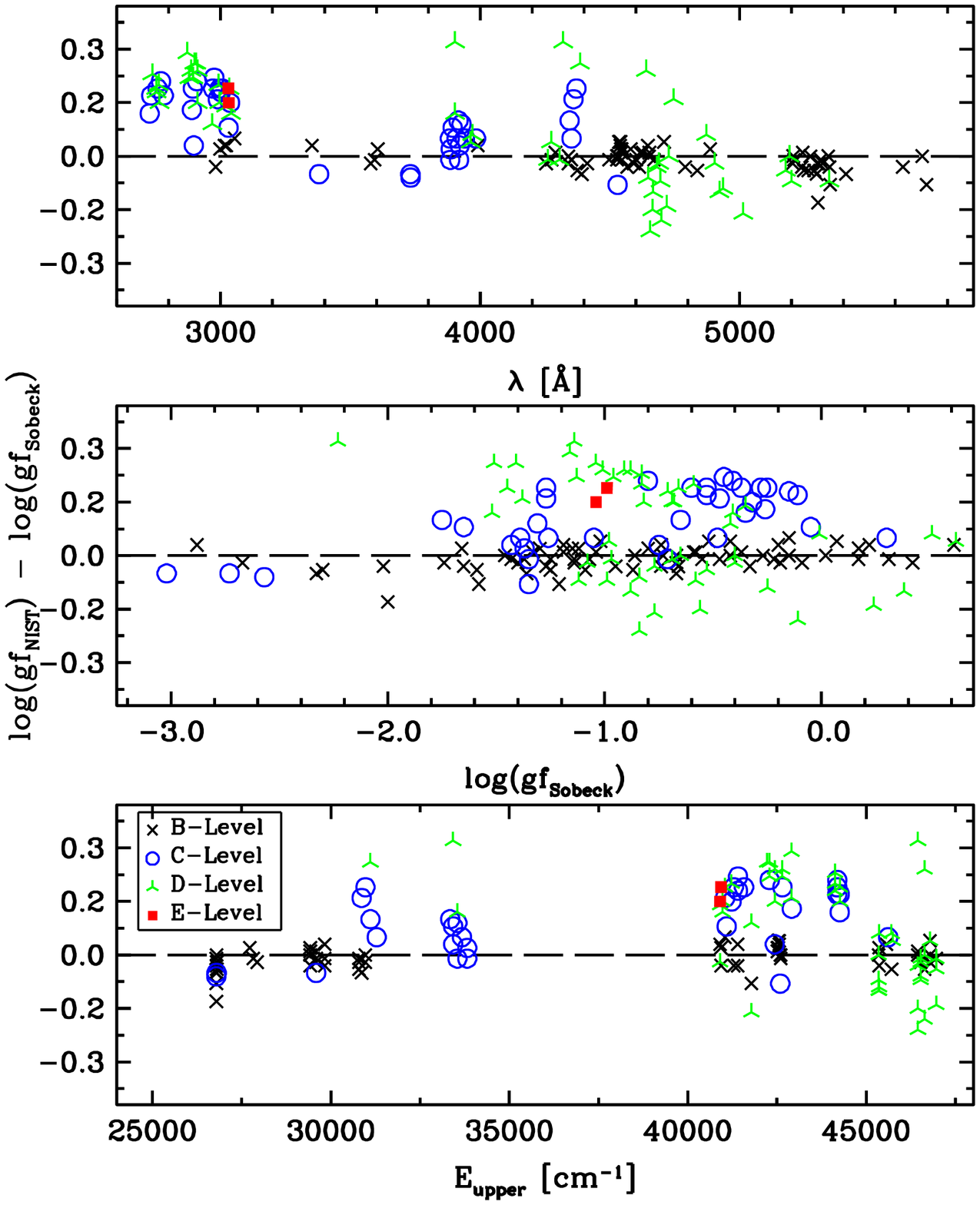}
\vspace*{0.2in}
\caption{Comparison of our oscillator strength values to those of the NIST compilation.  The upper panel shows the difference
         between the log($gf_{NIST}$) and log($gf_{Sobeck}$) as a function of wavelength.  
         The middle panel displays the difference verus the log($gf_{Sobeck}$) values.
         The bottom panel illustrates the difference as a function of upper energy level (E$_{upper}$). 
\label{f3}}
\end{figure}

NIST has assembled a collection of Cr I transition 
probability data from 11 different sources (Martin \etal\ 1988\nocite{Mar88}).  Classification of oscillator
strengths in terms of accuracy is done by NIST as follows: $A \leq 3\%$, $B \leq 10\%$, $C \leq 25\%$, $D \leq 50\%$,
and $E > 50\%$.  Figure~\ref{f3} presents our data in comparison to the NIST compilation. We share 155 transitions
with the NIST compilation with the average and rms values of $log(gf)_{NIST} - log(gf)_{Sobeck}$ 
tabulated to be $+$0.05 and 0.12 respectively.  A specific breakdown of the average and rms
differences of $log(gf)_{NIST} - log(gf)_{Sobeck}$ with respect to the NIST oscillator strength categorization
is as follows: B-level, $-$0.01 and 0.033 respectively; C-level: 0.10 and 0.13 respectively; 
D-level, 0.08 and 0.17 respectively; and E-level 0.17 and 0.17 respectively.  The upper panel of Figure~\ref{f3}
shows a systematic trend with wavelength (as the wavelength decreases, the mean agreement between
the two data sets diminishes and the line-to-line scatter increases). We have gf data for 2 E-level and
49 D-level accuracy transitions which should be taken as an improvement and given preference over the NIST values. 

%%%%%%%%%%%%%%%%%%%%%%%%%%%%%%%%%%%%%%%%%%%%%%%%%%%%%
\section{Solar and Stellar Chromium Abundances}
%%%%%%%%%%%%%%%%%%%%%%%%%%%%%%%%%%%%%%%%%%%%%%%%%%%%%%

Our new Cr I oscillator strength data are now applied to the solar spectrum and a few other stars.  We 
have chosen stars of varying metallicity and evolutionary state:
HD 75732 ([Fe/H] $= +0.35$; an extremely metal-rich dwarf); HD 140283 ([Fe/H] = $-2.50$; a very metal-poor subgiant); and 
CS 22892-052 ([Fe/H] $= -3.10$; a well-characterized, low metallicity, r-process rich giant).
Our Cr abundance analysis was facilitated by the existence of numerous transitions in the 
visible wavelength range and the presence of both the neutral and first-ionized species.

%%%%%%%%%%%%%%%%%%%%%%%%%%%%%%%%%%%%%%%%%%%%%%%%%%%%%%%%%%%%%%%%%%%%%%%%%%%%%%%%%%%%%%%%%%%%%%%%
\subsection{Inclusion of Oscillator Strengths for Singly-Ionized Chromium from the FERRUM Project}
%%%%%%%%%%%%%%%%%%%%%%%%%%%%%%%%%%%%%%%%%%%%%%%%%%%%%%%%%%%%%%%%%%%%%%%%%%%%%%%%%%%%%%%%%%%%%%%%

The availability of new Cr II transition probabilities (Nilsson \etal\ 2006)
enables us to compare the solar abundance value from the Cr I lines to that from the Cr II lines.
Nilsson \etal\ (2006) and our group both employ the same technique which is the most-advanced
and broadly applicable for determining transition probabilities in complex spectra.
With the combination of LIF radiative lifetime determinations and FTS branching fraction measurements, 
they have generated a complete set of gf values for the 25
lowest odd-parity energy levels of Cr II.  Nilsson \etal\ give oscillator strengths for 119 Cr II transitions in the
wavelength range 2050-4850 \AA\ with an uncertainty of $\sim10-15$\%.

%%%%%%%%%%%%%%%%%%%%%%%%%%%%%%%%%%%%%%%%%%%%%%%%%%%%%
\subsection{Line Selection and Analysis}
%%%%%%%%%%%%%%%%%%%%%%%%%%%%%%%%%%%%%%%%%%%%%%%%%%%%%%

Development of a line list suitable for stellar abundance analysis involved two selection criteria: detection of blends
and determination of relative line strength.  To review the numerous available chromium transitions,
we employed the solar spectral identification atlas of Moore et al. (1966\nocite{Moo66}) and the atomic line and 
parameter compendium VALD (Vienna Atomic Line Database, Kupka et al. 1999\nocite{Kup99}).  
We eliminated all lines with strong core blends and those with significant wing contaminants 
(within 0.1 \AA\ of the transition center).  As a result, a significant number of lines (slightly less than
40\%\ of the initial list) still remain at hand for abundance determinations.

Equivalent width analyses were sufficient to determine the elemental abundance of chromium as effects due to hyperfine and 
isotopic splitting were negligible.  Chromium has three stable isotopes: \iso{52}{Cr} $(83.79\% )$, \iso{53}{Cr} $(9.50\% )$, and  
\iso{54}{Cr} $(2.37\% )$.  A fourth isotope.\iso{50}{Cr} $(4.36\% )$, is metastable with an extremely 
long half-life ($\tau > 1.8E17$ years). The odd isotope \iso{53}{Cr} does not posses significant hyperfine structure 
as its nuclear g-factor is small (the dipole moment is $-0.47$ nuclear magnetons with $I = 3/2$).
The isotopic splitting of Cr lines is imperceptible (shifts are less than 0.007 \AA, Heilig \& Wendlandt 1967\nocite{Hei67}) 
and consequently, do not contribute to changes in abundance.  

To measure the equivalent widths (EWs), we used the interactive software package SPECTRE of Fitzpatrick \& Sneden (1987\nocite{Fit87}).  
For a particular transition, the numerical evaluation of the equivalent width was done via the fit of a 
Gaussian to the line profile.  We did not find any evidence of excessive damping in the wings of strong lines.  
A few synthetic spectrum computations were done as spot-checks and no appreciable 
gain in accuracy was found.  Table 4 lists the EW values for all target stars.

To gauge line strengths, we employed reduced widths, defined as $RW = EW/\lambda$.  
The evaluation of log (RW) for an individual line determines its position
on the curve of growth (COG).  Line saturation of (some) Cr I and Cr II transitions
is an issue and must be dealt with accordingly.  In the case of the Sun, transitions with a log (RW)$> -4.3$    
were found to lie on the exponential portion of the COG (consequently, insensitive to changes in abundance)
and were immediately discarded.

\begin{figure}
\epsscale{0.80}
\plotone{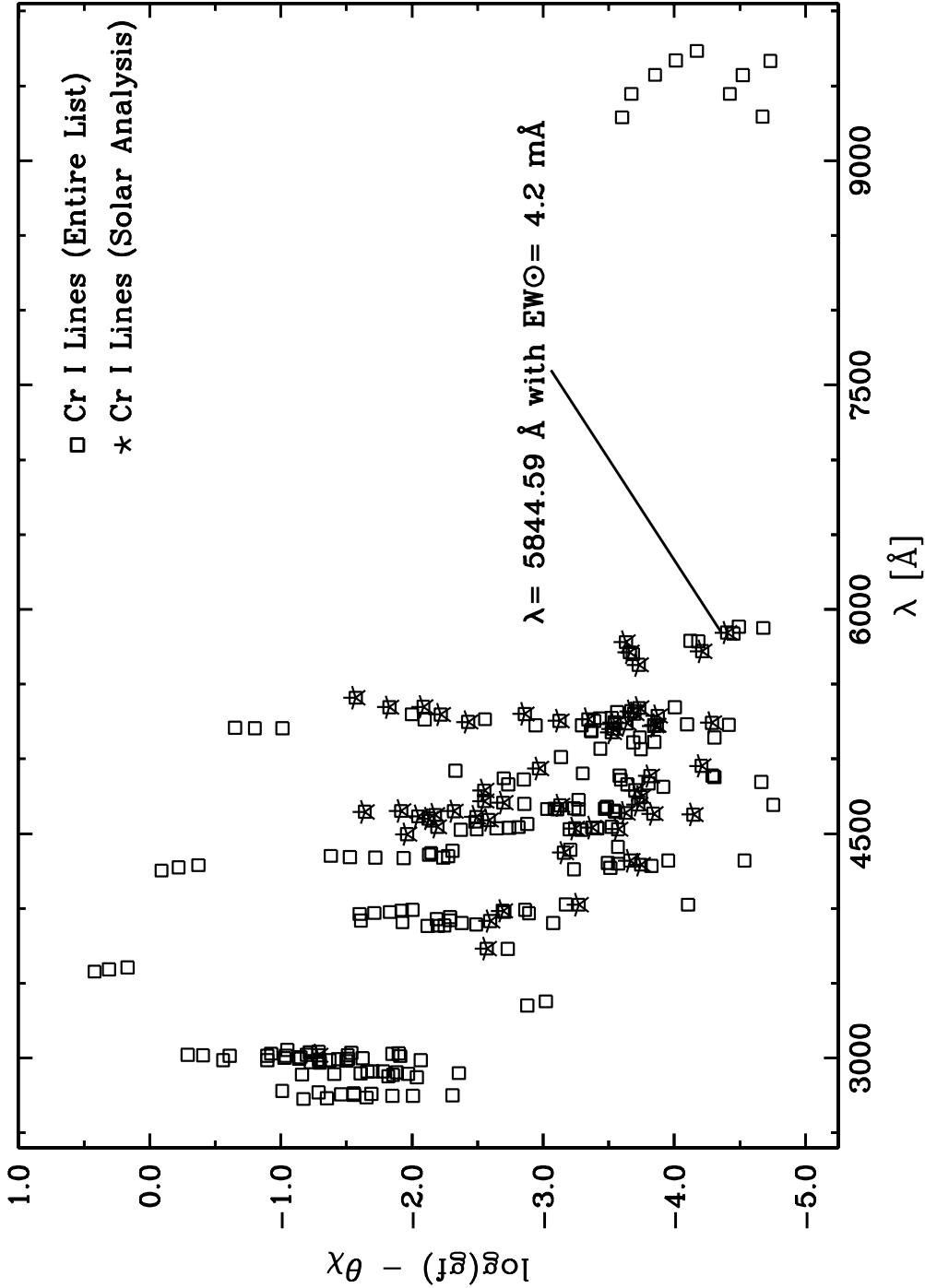}
\vspace*{0.2in}
\caption{Relative strength factors (RSF) as defined by $log(gf) - \theta\chi$ for the Cr I transitions.  Reduced widths of weak lines
         should be proportional to these factors.  For these computations, $\theta = 0.87$, the inverse of the effective temperature
         of the Sun.  The squares indicate the RSF for all 263 Cr I lines and the stars designate those Cr I lines actually used in 
         the derivation of the solar abundance.  The 5844.59 \AA\ line is specially noted in the plot as it is has a small RSF 
         yet is still detectable in the solar spectrum. 
\label{f4}}
\end{figure}

For an individual specie, the determination of stellar abundances under the stipulation of LTE  
allows for the relation of line strengths to both transition probabilities
and Boltzmann-Saha factors.  We may define a relative strength factor (RSF) as log $gf-\theta\chi$ where 
$\chi$ is the excitation energy in units of eV and $\theta$ is the standard inverse temperature
relation, $\theta = 5040/T$. Consequently for weak lines, log (RW) is directly proportional to
the RSF.  Figure~\ref{f4} shows our computation of the RSF for neutral chromium lines in 
the Sun (in which case $\theta$ becomes 0.87).  Notice that the 5844.59 \AA\ line with an extremely low RSF of    
$-$6.19 is still detectable in the Sun and that the strongest transitions are located in the ultraviolet and
blue visible portions of the spectrum.

With the aid of energy level information from the NIST database, the re-computation of the partition functions for both Cr I and Cr II
was done.\footnote{The relevant NIST website is: \url{http://physics.nist.gov/PhysRefData/ASD/levels\_form.html.}}  These data
were then compared to the partition functions from Irwin (1981\nocite{Irw81}) and to those from Halenka \& Grabowski (1986\nocite{Hal86}).  
Good agreement among the data sets was established and the newly computed partition functions for Cr II were used in the abundance
determination.

%%%%%%%%%%%%%%%%%%%%%%%%%%%%%%%%%%%%%%%%%%%%%%%%%%%%%%
\subsection{The Solar Photospheric Chromium Abundance}
%%%%%%%%%%%%%%%%%%%%%%%%%%%%%%%%%%%%%%%%%%%%%%%%%%%%%%

For cooler stars, main sequence stars (of high surface gravity) such as the Sun, 
collisional line broadening must be taken into account.  
In these types of stellar atmospheres, the broadening of strong spectral features is predominantly due to collisions 
with neutral hydrogen atoms.  The classical treatment of collisional broadening involves van der Waals theory and the
determination of the interaction energy parameter.  In the Unsold approximation (1927, 1955\nocite{Uns27} \nocite{Uns55}), the interaction energy
is related to the fixed energy debt, E$_{p}$, which is set to E$_{p} = 4/9$ AU for all atoms regardless of species 
(note that from this E$_{p}$, the familiar van der Waals damping coefficient, C$_{6}$ is obtained).     
An improved approach (applicable to both neutral and ionized species) was developed by Anstee, Barklem, \& O'Mara
(hereafter labeled ABO theory; Barklem \& Aspelund-Johasnsson 2005\nocite{Bar05}; Barklem et al. 1998\nocite{Bar98} and references therein).  
Essentially, it derives the interaction energy via the analytical determination to within a single numerical integration 
over the radial wavefunction of the perturbed atom (O'Mara \& Barklem 2003\nocite{Oma03}).  ABO theory calculations have been done
for tens of thousands of lines in the wavelength range 2300-13000 \AA\ of elements Li to Ni with a 
maximum error of about 20\%\ (Barklem \& Aspelund-Johansson 2005\nocite{Bar05}; Barklem et al. 2000\nocite{Bar00}).  
We made use of the publicly available collision damping constants (through the VALD catalog) 
and requested new calculations for those Cr I and Cr II transitions without published values (P. Barklem, priv. comm.).

The current version of the LTE line analysis code MOOG (Sneden 1973\nocite{Sne73}) was employed to calculate the abundances.
Our source of observed solar photospheric spectra was the center-of-disk spectral atlas of Delbouille et al. (1973).
Initially, we selected a Holweger \&\ M\"{u}ller (1974\nocite{Hol74}) model with a microturbulent velocity of \vt = 0.80 \kmsec.
Other model types we then tried include: MARCS (Gustaffson et al. 1975\nocite{Gus75}), ATLAS (Kurucz 1993\nocite{Kur93}),
Grevesse \& Sauval (1999\nocite{Gre99}), newMARCS (Gustaffson et al. 2003\nocite{Gus03}), and Asplund \etal\ (2004\nocite{Asp04}). 
Table 5 lists the abundance data from the various models.  We adopted the Holweger-M\"{u}ller model 
as it resulted in the smallest Cr I/II abundance difference as well as the lowest internal line scatter.  
The mean solar photospheric abundance for 58 lines of Cr I is
log$\epsilon(Cr) = 5.64 \pm 0.01$ ($\sigma = 0.07$) and 10 lines of Cr II is log$\epsilon(Cr) = 5.77 \pm 0.03$ ($\sigma = 0.13$).

\begin{figure}
\epsscale{0.80}
\plotone{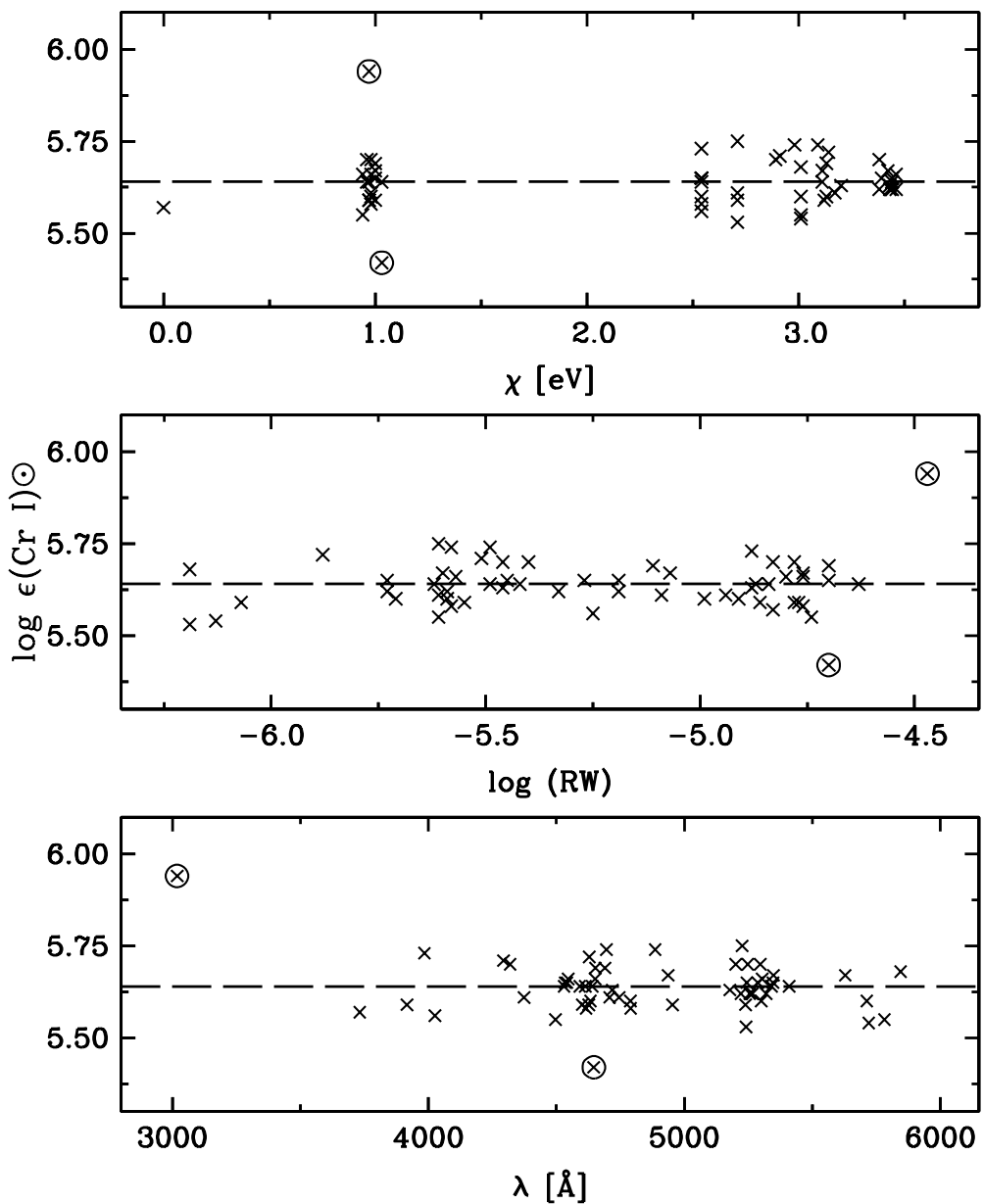}
\vspace*{0.2in}
\caption{Plot of solar Cr I abundances as a function of excitation potential ($\chi$), reduced width (log (RW)), and wavelength 
         ($\lambda$).  Encircled in each of the three panels are the two most erroneous abundance values.  
         Note that these two abundance data points correspond to lines that originate from major branches.  
         Consequently, the error in these two points cannot be attributed to oscillator
         strength uncertainties.
\label{f5}}
\end{figure}

Figure~\ref{f5} demonstrates that Cr I abundances do not exhibit any trends with equivalent width, excitation potential, or
wavelength.  In this figure, we have encircled the two most anomalous data points at 3018.49 \AA\ and 4646.15 \AA.  
Spectral line synthesis of these two transitions did not significantly change their respective abundance values.
The presence of unknown blends and the continuum determination are most certainly issues for the 3018.49 \AA\ line.
As for the feature at 4646.15 \AA\, we detect a slight line asymmetry, however we are not able to identify the
exact cause for its aberrant abundance (as the line originates from a dominant branch, has a highly accurate 
transition probability, and possesses no strong contaminants).  The result of the exclusion of these two transitions
from the abundance determination is log$\epsilon(Cr) = 5.64 \pm 0.01$ with $\sigma = 0.05$ (a slight decrease in the
standard deviation).

For error estimation, we consider the dependence of the chromium abundances on stellar atmospheric parameters and damping 
constants.  If we vary the \vt\ by $+0.2$/$-0.2$ \kmsec, we find that the Cr I abundance changes by $-0.04$/$+0.03$ dex and the Cr II
abundance by $-0.07$/$+0.06$ dex.  An alteration in \teff\ of $+100$/$-100$ K results in abundance changes in Cr I and Cr II of $+0.08$/$-0.07$ 
dex and $+0.00$/$-0.00$  dex respectively. A surface gravity variation of $\Delta\logg = +0.20/-0.20$ yields a $-0.02$/$0.0$ dex 
Cr I abundance change and a $+0.03$/$-0.04$ dex Cr II abundance change.  If we then employ 
a damping constant formulation as suggested by Blackwell et al. (1984b\nocite{Bla84b}; also mentioned in Simmons \& Blackwell 1982\nocite{Sim82})
as opposed to the Barklem values, our Cr I abundance decreases just slightly by 0.02 dex to $5.62\pm0.01$ 
and our Cr II abundance becomes $5.74\pm0.03$ (0.03 dex decrease).  It is apparent that the 
singly-ionized chromium abundance is more sensitive to these parameter adjustments (in the solar photosphere, 
Cr II is the dominant species as chromium has a relatively small first-ionization potential, 6.766 eV Grigoriev \& Meilikhov 1997\nocite{Gri97}).  
For stars with atmospheric parameters similar to the Sun, the total systematic error of the
abundance values is estimated to be 0.09 dex for both Cr I (largely attributed to uncertainty in \teff) and Cr II (mostly due to
uncertainty in \vt).  

One of the first studies to have had the benefit of both high quality spectra and transition probability data, 
Bi\'{e}mont et al. (1978\nocite{Bie78}) derived $5.64\pm0.10$ for the solar photospheric chromium abundance. The critical compilation of 
solar system abundances by Anders \& Grevesse (1989\nocite{And89}) recommends log$\epsilon(Cr I)_{\sun} = 5.67\pm0.03$ (subsequent 
publications, Grevesse et al. 1996\nocite{Gre96} and Grevesse \& Sauval 1998\nocite{Gre98}, restate this value).  Asplund et al. (2005a\nocite{Asp05a}
find log$\epsilon(Cr I)_{\sun} = 5.64\pm0.10$.  All of these numbers are in excellent agreement with the 
current meteoritic value of $5.63\pm0.05$ (Lodders 2003\nocite{Lod03}).  None of the solar abundance determinations use 
lines from singly-ionized chromium.  With 58 transitions, we have been able to derive a 
value for the solar abundance of neutral chromium, $5.64 \pm 0.01$, which is in good agreement with these values from literature. 

%%%%%%%%%%%%%%%%%%%%%%%%%%%%%%%%%%%%%%%%%%%%%%%%%%%%%%%%%%%%%%%%%%%%%%%%%%%%%%%%%%%%
\subsection{Detection of non-LTE Effects in Excitation for Neutral Chromium?}
%%%%%%%%%%%%%%%%%%%%%%%%%%%%%%%%%%%%%%%%%%%%%%%%%%%%%%%%%%%%%%%%%%%%%%%%%%%%%%%%%%%%

Blackwell \etal\ (1987\nocite{Bla87}) reported indications of non-LTE effects in excitation for lines of neutral chromium. 
The abundance derivation procedure of the Blackwell group is summarized as follows: use of the solar photospheric spectral atlas
by Delbouille et al. (1973\nocite{Del73}); determination of collisional damping constants; employment of both the 
Holweger-M\"{u}ller (1974\nocite{Hol74}) and MARCS (Gustaffson et al. 1975\nocite{Gus75}) model atmospheres; and EW measurement via 
a synthetic line profile fit to an observed transition.  In contrast to the Cr I lines of higher excitation potential, 
they found a noticeably larger spread in the abundances from the 1 eV lines. Particularly for three transitions of the z$^{5}$P
multiplet (5247.57 \AA, 5300.75 \AA, and 5345.80 \AA), Blackwell et al. reported a markedly higher abundance 
($<log\epsilon_{CrI-z^5P}> = 5.81$ as opposed to $<log\epsilon_{CrI}> = 5.69$).  They did not
consider that the oscillator strengths of these 1 eV lines as sources of major error since they agreed well with those gf values
given by Tozzi et al. (1985).  Nor did they believe the equivalent widths were at fault (in the case of the z$^{5}$P multiplet) as
Blackwell et al. were not able to detect any blends.  On the basis of these two pieces of evidence, the Blackwell group 
concluded that non-LTE did indeed affect these Cr I lines.

In the present study, we re-examine these low excitation chromium
transitions.  We do not find an abnormally large scatter in the abundances of the 1 eV lines (in fact, the
standard deviation for these lines was $\sigma = 0.06$; exactly the same as that for the entire line list). Also,
the three transitions of the z$^{5}$P multiplet do not appear to give an anomalously high abundance    
($log\epsilon_{CrI-z^5P} = 5.65$ as compared to $<log\epsilon_{CrI}> = 5.64$).  In addition to these three lines,
we derive the abundances for 3 more transitions of the z$^{5}$P multiplet with the average for all lines
equal to 5.70.  Table 6 displays all of the results and shows that our transition probabilities for these lines
agree very well with those of Blackwell et al.  This table also shows a comparison of three sets of EW measurements. 
We see that our values agree very well with those of Moore et al. (1966\nocite{Moo66}) whereas the Blackwell EWs report consistently higher 
than ours.   We are not able to pinpoint the exact reasons for the discrepancy between the Blackwell et al. data 
and our own for these z$^{5}$P transitions.  In summary, we do not find any compelling evidence for departures from LTE
in the transitions of Cr I. 

%%%%%%%%%%%%%%%%%%%%%%%%%%%%%%%%%%%%%%%%%%%%%%%%%%%%%
\subsection{Chromium Abundances in Other Stars}
%%%%%%%%%%%%%%%%%%%%%%%%%%%%%%%%%%%%%%%%%%%%%%%%%%%%%%

We now consider how the new Cr I and Cr II transition probability data affects the [Cr/Fe] ratios in other stars.  
In a preliminary investigation, we have derived new values in three stars with previously established chromium abundances.  
These stars represent extremes in metallicity and exhibit different evolutionary states.  We modified the initial line 
list to account for the unique blending and detectability concerns of each star and performed EW measurements.  Table
4 gives the EW data for all of the stars.  We used an interpolation software program (kindly
provided by I. Ivans and A. McWilliam) to generate model atmospheres from the ATLAS grid (Kurucz 1993\nocite{Kur93}).  
We then proceeded with abundance determinations in the manner described above.

HD 75732 ($\rho^{\rm 1}$ Cnc) is a metal-rich disk main sequence star which is host to a planetary system 
that was first detected by Butler \etal\ (1997\nocite{But97}).  The most recent publication of chromium
abundances for this star is from the large survey by Luck \& Heiter (2005\nocite{Luc05}).  We adopted the model atmosphere 
parameters for HD 75732 as reported by Valenti \& Fischer (2005\nocite{Val05}): ($T_{eff}$/log$g$/[Fe/H]) = (5235/4.45/+0.25)
which differ from those listed by Luck \& Heiter: ($T_{eff}$/log$g$/[Fe/H]/$v_t$) = (5375/4.35/+0.50/0.45).
We derived abundances of log$\epsilon$(\ion{Cr}{1})~= 5.98 ($\sigma = 0.12$, 31 lines) 
and log$\epsilon$(\ion{Cr}{2})~= 6.22 ($\sigma = 0.05$, 3 lines).  The Luck \& Heiter (2005) values
for this star $log\epsilon(CrI) = 6.15$ ($\sigma = 0.14$) and $log\epsilon(CrII) = 6.25$ ($\sigma = 0.12$) are 
comparable to ours (the difference between the values falls within the stated uncertainties).  The discordance
between the abundances from \ion{Cr}{1} and \ion{Cr}{2} lines for HD 75732 is 0.24 dex.  With the 
use of the \S4.3 solar abundances, we find that [Cr/H]$_{\rm I} = +0.34$ and [Cr/H]$_{\rm II} = +0.45$
(which further confirms the super-metal-rich status of this star).

The subgiant HD 140283 was one of the first very metal-poor stars to be discovered (Chamberlain \& Aller 1951\nocite{Cha51})
and has been well-studied over the past several decades.  The Cr I abundance for this star was 
reported by King et al. (1998\nocite{Kin98}): log~$\epsilon$(\ion{Cr}{1}) = 2.85 ($\sigma = 0.10$).  With the model atmospheric parameters 
suggested by I. Ivans (5725/3.65/-2.20/1.10; 2006 priv.comm.), we obtained log~$\epsilon$(\ion{Cr}{1})~= 2.86 
($\sigma =0.04$, 13 lines) and log~$\epsilon$(\ion{Cr}{2})~= 3.16 ($\sigma = 0.14$, 11 lines).  The King \etal\ value
agrees well with ours though it is based on a single line.  For HD 140283, we also determined the 
differential abundances [Cr/H]$_{\rm I} = -2.83$ and [Cr/H]$_{\rm II} = -2.65$.

Sneden et al. (1994\nocite{Sne94}, 2003\nocite{Sne03}) detected a significant enhancement of $r$-process neutron-capture 
elements in the very metal-poor giant CS~22892-052. They were also able to determine the
chromium abundances for this star: log~$\epsilon$(\ion{Cr}{1})~= 2.33 ($\sigma = 0.11$, 6 lines) and
log~$\epsilon$(\ion{Cr}{2})~= 2.42 ($\sigma = 0.14$, 2 lines).  The model atmospheric
parameters (4800/1.50/--3.12/1.95) from Sneden \etal\ 2003\nocite{Sne03} were used to derive: 
log~$\epsilon$(\ion{Cr}{1})~= 2.31 ($\sigma = 0.13$, 9 lines) and log~$\epsilon$(\ion{Cr}{2})~= 2.54 ($\sigma = 0.13$, 7 lines).  
Our values agree reasonably well with those reported by Sneden et al (and we believe supersede them).  
In addition, we find for CS~22892-052 the differential abundances: [Cr/H]$_{\rm I} = -3.33$ and [Cr/H]$_{\rm II} = -3.23$.

These data offer a brief snapshot of the chromium abundance trend with metallicity in the Galaxy.
They suggest that the disagreement in abundance values from Cr I and Cr II widens as the metallicity decreases
(the derivation of Fe abundances and subsequent [Cr/Fe] determination awaits an investigation with
a larger available data pool analyzed in a consistent manner).  The difference appears to grow 
from $\simeq -0.1$ at [Fe/H]~$>$~0 to perhaps as much as $\simeq -0.3$ at [Fe/H]~$< -2.5$ 
(though the effect is substantially lessened if the solar abundance discrepancy between the two species 
is acknowledged). Finally, we emphasize that the chosen model parameters are taken from the literature, 
and these choices impact the derived abundances.

%%%%%%%%%%%%%%%%%%%%%%%%%%%%%%%%%%%%%%%%%%%%%%%%%%%%%%%%%%%%%%%%%%%%%%%%%%%%%%%%%%%%
\subsection{Chromium Ionization Imbalance Result of Departures from LTE?}
%%%%%%%%%%%%%%%%%%%%%%%%%%%%%%%%%%%%%%%%%%%%%%%%%%%%%%%%%%%%%%%%%%%%%%%%%%%%%%%%%%%%

We have shown that factors such as model grid selection, stellar atmospheric
parameter choice, and equivalent width measurement technique cannot fully account for
the sizable abundance discrepancy between the Cr I and Cr II lines (which was detected in all stars). 
In the case of the Sun, the difference between $log\epsilon_{CrI} = 5.64$ ($\sigma = 0.07$)
and $log\epsilon_{CrII} = 5.77$ ($\sigma = 0.13$) is $\Delta = 0.13$.  We note though that this difference does 
fall within the error of $\sigma = 0.15$ (by the quadrature addition of sigmas).

The inability to reconcile the abundances from the neutral and first-ionized states of a particular species is not unique.
Even with the reference element Fe there are difficulties: the solar abundances from Fe I and Fe II transitions are likewise discordant.
In the 1990's, multiple papers quoted different values for the photospheric abundance of Fe I and Fe II (\eg\
Milford \etal\ 1994\nocite{Mil94}; Blackwell \etal\ 1995\nocite{Bla95}; Holweger \etal\ 1995\nocite{Hol95}; 
Anstee \etal\ 1997\nocite{Ans97}; and Schnabel \etal\ 1999\nocite{Sch99}).  We are not aware of a study that 
simultaneously examines both Fe I and Fe II with atomic data obtained from one, sole 
laboratory technique and with abundances derived from a single, consistent methodology.
   
For every element, accurate determinations of the oscillator strengths for the majority ionic species is critical.
Numerous challenges face laboratory spectroscopists.  Large wavelength separations between
dominant UV branches (2000 \AA\ to 3000 \AA) and minor IR branches (7000 \AA\ to 10000 \AA) from the same upper level hinder the 
radiometric calibration process (the frequency cubed scaling of the A values means for instance that a 2500 \AA\ branch is 27$\times$ stronger than 
7500 \AA\ branch with a similar dipole matrix element from the same upper level). 
For the dominant UV branches, the uncertainty depends upon the lifetime measurement.  However, it is rarely useful to obtain 
abundances from these branches as they generally correspond to highly-saturated transitions.  
On the other hand for the weak IR branches, the uncertainty limitation is the
branching fractions (as by definition they must sum to 1.0).  Errors may also result from
optical depth effects.  The possibility exists for the weaker branches to drop into the noise before
the discharge current is low enough to ensure that the dominant UV branches are optically thin.

\begin{figure}
\epsscale{0.80}
\plotone{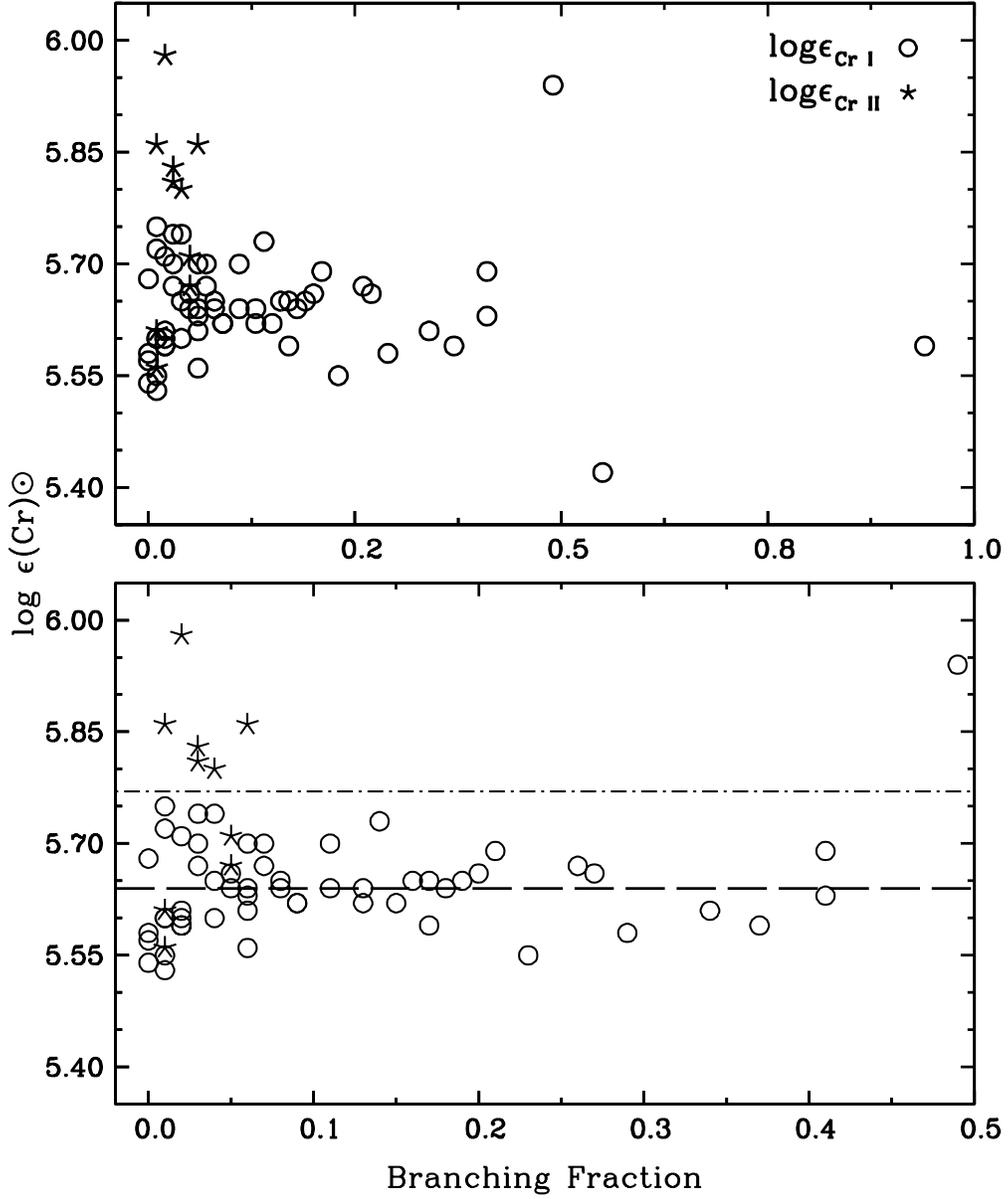}
\vspace*{0.2in}
\caption{Branching Fractions for the Cr I and Cr II lines used in the solar abundance analysis.  The lower
         panel (which is an enlarged view of the upper) shows the average abundance from the Cr I transitions (dashed line) 
         as well as that from the Cr II transitions (dash-dotted line). 
\label{f6}}
\end{figure}

In spite of these cautions, we do not believe that the transition probabilities are a significant
error source.  We took all possible steps to ensure
the rigorous determination of the Cr I gf-values.  The Cr II oscillator strengths from the FERRUM Project 
(Nilsson \etal\ 2006\nocite{Nil06})
are of the highest quality (``state-of-the-art'' techniques were employed).  They took steps
to avoid optical depth effects.  Furthermore, the Cr I/Cr II abundance disagreement 
is not consistent with optical depth problems in the Cr II branching
fraction study.  Note that the Nilsson \etal\ branching fractions compare well
with those generated by the Cowan code (1981\nocite{Cow81}).  Figure~\ref{f6} displays the branching fractions
associated with the Cr I and Cr II transitions used in the solar abundance analysis.  As shown,
only the weak branches of Cr II (with inherent errors in their associated A values of 9-37\%) were employed.

Given that the average abundance from Cr I lines is markedly lower
than that from Cr II lines and the ostensible reliability of the
oscillator strengths, we should now consider other possible causes with special 
focus on departures from LTE.  Unfortunately, we have not been able to locate
any published set of non-LTE calculations for Cr.  Eventually, non-LTE effects will be quantified for a variety of elements  
(Asplund \etal\ 2005b\nocite{Asp05b} and references therein).  It is reasonable to expect that the non-LTE effects 
on the ionization balance will be larger than non-LTE effects on level populations in a single ionization stage.  
Estimations of the non-LTE influence on the solar photospheric abundances of some elements have been done by several groups.
For example, Shchukina \& Trujillo (2001\nocite{Sch01}) suggest that non-LTE effects for Fe I lines might be as large as 0.1 dex. 
Future work on chromium should include (for instance) the precise re-measurement of the minor Cr II branches 
and the commencement of statistical equilibrium calculations.

%%%%%%%%%%%%%%%%%%%%%%%%%%%%%%%%%%%%%%%%%%%%%%%%%%%%%
\section{Conclusion}
%%%%%%%%%%%%%%%%%%%%%%%%%%%%%%%%%%%%%%%%%%%%%%%%%%%%%%

Published lifetimes combined with branching fractions measured with Fourier transform spectrometry to determine
transition probabilities for 263 lines of Cr I.  This improved set of oscillator strengths
has been used to determine the solar photospheric abundance of Cr I, log$\epsilon = 5.64 \pm 0.01$ ($\sigma = 0.07$), 
from 58 lines.  The spectra of three other stars (HD 140283, CS 22892-052, and HD 75732) was analyzed, employing 9 to 31 
Cr I lines per star. 

With the use of plane-parallel models and the assumption of LTE conditions, we were not able to achieve
ionization equilibrium in chromium.  Abundances from Cr I transitions were consistently underabundant with respect to 
those from Cr II lines. We have speculated as to the possible causes of the discrepancy.  
Contributions from internal and external error sources cannot account
for the difference.  We note that the suppression of the Cr I abundance relative to that of Cr II
is commensurate with the idea of Cr I overionization.  Therefore, we believe that the discrepancy may be due to non-LTE
effects.  Our contention is not novel in that several other groups have suggested that departures from LTE effect 
element abundances in the Sun (\eg\ Shchukina \& Trujillo Beno 2001\nocite{Sch01}, Takeda et al. 2005\nocite{Tak05}).

Steps toward the resolution of the chromium ionization imbalance problem include the re-measurement of the Cr II
branching fractions and the reanalysis of the Cr abundance with a 3-dimensional hydrodynamical model.

\acknowledgments
We thank Inese Ivans for the use of her Keck I HIRES HD 140283 spectra and Carlos Allende Prieto for the
use of his HD 75723 spectra (from his S$^{4}$N solar neighborhood stars database).  We are deeply indebted to Paul Barklem for 
providing Cr I and Cr II damping constants.  We are grateful to the NSF 
through grants AST 05-06324 to J. L. and AST 03-07495 to C. S. for providing funding support for this research.

\newpage

\newpage

\newpage

\newpage

\newpage

\newpage

\tablenum{1}
\tablecolumns{7}
\tablewidth{0pt}

\begin{deluxetable}{lccccccc}
\tablecaption{Radiative Lifetimes of 65 Cr I Levels from LIF Measurements}

\tablehead{
\colhead{Configuration}                     &
\colhead{Term}                              &
\colhead{J}                                 &                           
\colhead{Level}                             &
\colhead{$\tau$ [ns]}                       &
\colhead{$\tau$ [ns]}                       &
\colhead{$\tau$ [ns]}                       \\
\colhead{}                                  &
\colhead{}                                  &
\colhead{}                                  &
\colhead{[cm$^{-1}]$}                       &
\colhead{Cooper et al. 1997}                &
\colhead{Other LIF\tablenotemark{a}}        &
\colhead{Martin et al. 1988}                \\
}
\startdata3d$^{5}$($^{6}$S)4p	&	z $^{7}$P$^{o}$	&	2	&	23305.01	&	32.2	$\pm$	1.6	&	31.42	$\pm$	0.25	$^{a}$	&	31.6	\\
	&		&		&		&				&	31.2	$\pm$	1.0	$^{b}$	&	\nodata	\\
	&		&	3	&	23386.35	&	31.5	$\pm$	1.6	&	32.22	$\pm$	0.17	$^{a}$	&	32.5	\\
	&		&	4	&	23498.84	&	30.3	$\pm$	1.5	&	31.15	$\pm$	0.08	$^{a}$	&	31.7	\\
	&		&		&		&				&	31.8	$\pm$	2.5	$^{c}$	&	\nodata	\\
	&		&		&		&				&	31.6	$\pm$	0.5	$^{e}$	&	\nodata	\\
3d$^{5}$($^{6}$S)4p	&	z $^{5}$P$^{o}$	&	3	&	26787.50	&	16.2	$\pm$	0.8	&	17.5	$\pm$	0.9	$^{d}$	&	17.0	\\
	&		&	2	&	26796.28	&	16.2	$\pm$	0.8	&	16.7	$\pm$	0.8	$^{d}$	&	16.5	\\
	&		&	1	&	26801.93	&	16.0	$\pm$	0.8	&	17.3	$\pm$	0.9	$^{d}$	&	16.7	\\
3d$^{4}$($^{5}$D)4s4p($^{3}$P$^{o}$)	&	y $^{7}$P$^{o}$	&	2	&	27728.87	&	6.6	$\pm$	0.3	&	\nodata				&	6.2	\\
	&		&	3	&	27820.23	&	6.6	$\pm$	0.3	&	\nodata				&	6.7	\\
	&		&	4	&	27935.26	&	6.6	$\pm$	0.3	&	\nodata				&	6.8	\\
3d$^{4}$($^{5}$D)4s4p($^{3}$P$^{o}$)	&	y $^{5}$P$^{o}$	&	1	&	29420.90	&	76.6	$\pm$	3.8	&	72.8	$\pm$	5.5	$^{d}$	&	75.8	\\
	&		&	2	&	29584.62	&	72.9	$\pm$	3.6	&	70.6	$\pm$	3.6	$^{d}$	&	72.5	\\
	&		&	3	&	29824.75	&	69.1	$\pm$	3.5	&	63.5	$\pm$	3.0	$^{d}$	&	66.5	\\
3d$^{4}$($^{5}$D)4s4p($^{3}$P$^{o}$)	&	z $^{5}$F$^{o}$	&	1	&	30787.30	&	101	$\pm$	5	&	\nodata				&	110	\\
	&		&	2	&	30858.82	&	99.1	$\pm$	5.0	&	\nodata				&	89.5	\\
	&		&	3	&	30965.46	&	99.9	$\pm$	5.0	&	\nodata				&	89.9	\\
	&		&	4	&	31106.37	&	94.5	$\pm$	4.7	&	\nodata				&	73.0	\\
	&		&	5	&	31280.35	&	91.3	$\pm$	4.6	&	\nodata				&	83.3	\\
3d$^{4}$($^{5}$D)4s4p($^{3}$P$^{o}$)	&	z $^{5}$D$^{o}$	&	0	&	33338.20	&	122	$\pm$	6	&	\nodata				&	103	\\
	&		&	1	&	33423.79	&	102	$\pm$	5	&	\nodata				&	87.0	\\
	&		&	2	&	33542.11	&	88.7	$\pm$	4.4	&	\nodata				&	79.4	\\
	&		&	3	&	33671.55	&	83.9	$\pm$	4.2	&	\nodata				&	77.5	\\
	&		&	4	&	33816.06	&	83.7	$\pm$	4.2	&	\nodata				&	87.7	\\
3d$^{4}$($^{5}$D)4s4p($^{1}$P$^{o}$)	&	y $^{5}$F$^{o}$	&	1	&	40906.46	&	3.4	$\pm$	0.2	&	\nodata				&	2.8	\\
	&		&	2	&	40971.29	&	4.5	$\pm$	0.2	&	\nodata				&	$<$ 8.6	\\
	&		&	3	&	41086.26	&	3.4	$\pm$	0.2	&	\nodata				&	$<$ 9.1	\\
	&		&	4	&	41224.78	&	3.4	$\pm$	0.2	&	\nodata				&	$<$ 18.5	\\
	&		&	5	&	41393.47	&	3.5	$\pm$	0.2	&	\nodata				&	3.3	\\
3d$^{4}$($^{5}$D)4s4p($^{1}$P$^{o}$)	&	x$^{5}$P$^{o}$	&	1	&	40930.31	&	5.6	$\pm$	0.3	&	\nodata				&	$<$ 12.0	\\
	&		&	3	&	41043.35	&	6.1	$\pm$	0.3	&	\nodata				&	$<$ 6.0	\\
3d$^{4}$($^{5}$D)4s4p($^{1}$P$^{o}$)	&	y $^{5}$D$^{o}$	&	0	&	41224.80	&	5.0	$\pm$	0.3	&	\nodata				&	3.3	\\
	&		&	1	&	41289.17	&	4.8	$\pm$	0.2	&	\nodata				&	$<$ 4.0	\\
	&		&	2	&	41409.03	&	4.7	$\pm$	0.2	&	\nodata				&	$<$ 4.0	\\
	&		&	3	&	41575.10	&	4.6	$\pm$	0.2	&	\nodata				&	$<$ 6.1	\\
	&		&	4	&	41782.19	&	4.5	$\pm$	0.2	&	\nodata				&	$<$ 23.3	\\
3d$^{4}$(a $^{3}$P)4s4p($^{3}$P$^{o}$)	&	x $^{5}$D$^{o}$	&	0	&	42218.37	&	13.4	$\pm$	0.7	&	\nodata				&	7.7	\\
	&		&	1	&	42292.96	&	13.7	$\pm$	0.7	&	\nodata				&	8.6	\\
	&		&	2	&	42438.82	&	14.5	$\pm$	0.7	&	\nodata				&	11.0	\\
	&		&	3	&	42648.26	&	16.0	$\pm$	0.8	&	\nodata				&	10.0	\\
	&		&	4	&	42908.57	&	17.6	$\pm$	0.9	&	\nodata				&	12.5	\\
3d$^{5}$($^{4}$G)4p	&	z $^{5}$G$^{o}$	&	2	&	42515.35	&	48.7	$\pm$	2.4	&	48.5	$\pm$	2.5	$^{d}$	&	49.3	\\
	&		&	3	&	42538.81	&	49.0	$\pm$	2.5	&	46.0	$\pm$	2.5	$^{d}$	&	46.3	\\
	&		&	4	&	42564.85	&	48.8	$\pm$	2.4	&	46.9	$\pm$	2.5	$^{d}$	&	47.2	\\
	&		&	5	&	42589.25	&	48.7	$\pm$	2.4	&	48.2	$\pm$	2.5	$^{d}$	&	48.8	\\
	&		&	6	&	42605.81	&	50.0	$\pm$	2.5	&	50.2	$\pm$	2.5	$^{d}$	&	51.0	\\
3d$^{5}$($^{6}$S)5p	&	w $^{5}$P$^{o}$	&	1	&	44125.90	&	5.5	$\pm$	0.3	&	\nodata				&	$<$ 4.3	\\
	&		&	2	&	44186.92	&	5.4	$\pm$	0.3	&	\nodata				&	$<$ 3.9	\\
	&		&	3	&	44259.36	&	5.2	$\pm$	0.3	&	\nodata				&	3.8	\\
3d$^{5}$($^{4}$G)4p	&	z 3H$^{o}$	&	6	&	45348.73	&	15.6	$\pm$	0.8	&	\nodata				&	20	\\
	&		&	5	&	45354.18	&	15.6	$\pm$	0.8	&	\nodata				&	$<$ 21.1	\\
	&		&	4	&	45358.63	&	15.5	$\pm$	0.8	&	\nodata				&	$<$ 16.7	\\
3d$^{5}$($^{4}$G)4p	&	y $^{5}$H$^{o}$	&	3	&	45566.02	&	8.8	$\pm$	0.4	&	\nodata				&	$<$ 9.3	\\
	&		&	4	&	45614.88	&	8.9	$\pm$	0.4	&	\nodata				&	$<$ 9.2	\\
	&		&	5	&	45663.28	&	8.9	$\pm$	0.4	&	\nodata				&	$<$ 357.	\\
	&		&	6	&	45707.36	&	8.8	$\pm$	0.4	&	\nodata				&	$<$ 8.2	\\
	&		&	7	&	45741.49	&	8.5	$\pm$	0.4	&	\nodata				&	7.7	\\
3d$^{4}$ 4s5s	&	f $^{7}$D	&	1	&	46448.60	&	8.7	$\pm$	0.4	&	9.2	$\pm$	0.7	$^{c}$	&	$<$ 12.0	\\
	&		&	2	&	46524.84	&	8.7	$\pm$	0.4	&	9.5	$\pm$	0.8	$^{c}$	&	$<$ 15.8	\\
	&		&	3	&	46637.21	&	8.6	$\pm$	0.4	&	9.2	$\pm$	0.7	$^{c}$	&	$<$ 15.4	\\
	&		&	4	&	46783.06	&	8.7	$\pm$	0.4	&	9.7	$\pm$	0.8	$^{c}$	&	$<$ 14.8	\\
	&		&	5	&	46958.98	&	8.7	$\pm$	0.4	&	9.8	$\pm$	0.8	$^{c}$	&	$<$ 17.6	\\
3d$^{4}$($^{3}$H)4s4p($^{3}$P$^{o}$)	&	x $^{5}$G$^{o}$	&	2	&	47047.47	&	16.3	$\pm$	0.8	&	15.0	$\pm$	1.5	$^{d}$	&	15.9	\\
	&		&	3	&	47125.70	&	16.9	$\pm$	0.8	&	15.2	$\pm$	1.0	$^{d}$	&	16.4	\\
	&		&	4	&	47189.87	&	16.0	$\pm$	0.8	&	\nodata				&	$<$ 476.	\\
	&		&	6	&	47222.27	&	13.2	$\pm$	0.7	&	13.9	$\pm$	0.7	$^{d}$	&	12.3	\\
	&		&	5	&	47228.80	&	14.9	$\pm$	0.7	&	14.8	$\pm$	0.7	$^{d}$	&	$<$ 455.	\\
\enddata
\tablenotetext{a}{LIF Literature References:
                 (a) Measures et al. 1977. (b) Kwong \& Measures 1980. (c) Marek 1975. (d) Kwiatowski et al. 1981. 
                 (e) Hannaford \& Lowe 1981.}
\end{deluxetable}

\begin{landscape}
\tablenum{2}
\tablecolumns{12}
\tablewidth{0pt}
\tabletypesize{\tiny}

\begin{deluxetable}{lccccccccccc}
\tablecaption{FTS Spectra Chosen for Branching Fraction Determination}

\tablehead{
\colhead{Spectrum}                          &
\colhead{Date}                              &                           
\colhead{Serial}                            &
\colhead{HC}                                &
\colhead{I$_{Discharge}$}                   &
\colhead{P$_{Buffer}$}                      &
\colhead{Number of}                         &
\colhead{Spectral Coverage}                 &
\colhead{Limit of}                          &
\colhead{Beam}                              &
\colhead{Filter(s)}                         &
\colhead{Photodiode}                        \\
\colhead{Number}                            &
\colhead{Recorded}                          &
\colhead{Number}                            &
\colhead{Discharge}                         &
\colhead{[Amps]}                            &
\colhead{[Torr]}                            &
\colhead{Co-adds}                           &
\colhead{[cm$^{-1}]$}                       &
\colhead{Resolution [cm$^{-1}]$}            &
\colhead{Splitter}                          &
\colhead{}                                  &
\colhead{Detector}                          \\
}
\startdata
~~~1 & 06-25-1982 & 7 & Cr-Ar & 0.50 & 0.65 & 8  & 7664-44591  & 0.057 & UV  &  CS 9-54       & Mid Range Si \\
~~~2 & 06-26-1982 & 4 & Cr-Ar & 0.10 & 1.00 & 6  & 7664-44591  & 0.057 & UV  &  CS 9-54       & Mid Range Si \\
~~~3 & 07-26-1984 & 6 & Cr-Ne & 0.75 & 3.00 & 4  & 7985-45407  & 0.054 & UV  &  WG295         & Mid Range Si \\
~~~4 & 07-26-1984 & 7 & Cr-Ne & 1.50 & 3.30 & 4  & 7985-45407  & 0.054 & UV  &  WG295         & Mid Range Si \\
~~~5 & 07-26-1984 & 8 & Cr-Ne & 1.50 & 3.30 & 4  & 7985-45407  & 0.054 & UV  &  WG295         & Mid Range Si \\
~~~6 & 02-28-1980 & 1 & Cr-Ar & 0.50 & 2.50 & 4  & 7908-28921  & 0.035 & Vis &  GG375         & Super Blue Si\\
~~~7 & 03-25-1980 & 3 & Cr-Ne & 0.95 & 4.00 & 10 & 13489-27089 & 0.034 & Vis &  GG400/CS 4.96 & Super Blue Si\\
\enddata
\end{deluxetable}
\end{landscape}

\newpage
\tablenum{3}
\tablecolumns{10}
\tablewidth{0pt}
\tabletypesize{\normalsize}

\begin{deluxetable}{lcccccccccccc}
\tablecaption{Atomic Transition Probabilities for Cr I Organized by Increasing Wavelength in $\lambda_{air}$}

\tablehead{
\colhead{$\lambda_{air}$}                   &
\colhead{E$_{upper}$}                       &
\colhead{Term}                              &
\colhead{J$_{upper}$}                       &
\colhead{E$_{lower}$}                       &
\colhead{Term}                              &
\colhead{J$_{lower}$}                       &
\colhead{A$_{LS calc}$}                     &
\colhead{A$_{Exp}$}                         &
\colhead{log gf}                           \\
\colhead{[\AA]}                             &
\colhead{[cm$^{-1}$]}                       &
\colhead{}                                  &
\colhead{}                                  &
\colhead{[cm$^{-1}$]}                       &
\colhead{}                                  &
\colhead{}                                  &
\colhead{[10$^{6}$s$^{-1}$]}                &
\colhead{[10$^{6}$s$^{-1}$]}                &
\colhead{}                                 \\
}
\startdata
2726.50	&	44259.36	&	w$^{5}$P$^{o}$	&	3	&	7593.16	&	a$^{5}$S	&	2	&	\nodata	&	58$\pm$3	&	-0.35	\\
2731.90	&	44186.92	&	w$^{5}$P$^{o}$	&	2	&	7593.16	&	a$^{5}$S	&	2	&	\nodata	&	52$\pm$3	&	-0.53	\\
2736.46	&	44125.90	&	w$^{5}$P$^{o}$	&	1	&	7593.16	&	a$^{5}$S	&	2	&	\nodata	&	43$\pm$4	&	-0.83	\\
2748.24	&	44186.92	&	w$^{5}$P$^{o}$	&	2	&	7810.82	&	a$^{5}$D	&	1	&	\nodata	&	12.3$\pm$2.0	&	-1.16	\\
2748.32	&	44125.90	&	w$^{5}$P$^{o}$	&	1	&	7750.78	&	a$^{5}$D	&	0	&	\nodata	&	29$\pm$3	&	-1.01	\\
2751.59	&	44259.36	&	w$^{5}$P$^{o}$	&	3	&	7927.47	&	a$^{5}$D	&	2	&	\nodata	&	4.42$\pm$0.26	&	-1.45	\\
2752.86	&	44125.90	&	w$^{5}$P$^{o}$	&	1	&	7810.82	&	a$^{5}$D	&	1	&	\nodata	&	57$\pm$4	&	-0.71	\\
2757.09	&	44186.92	&	w$^{5}$P$^{o}$	&	2	&	7927.47	&	a$^{5}$D	&	2	&	\nodata	&	44$\pm$3	&	-0.60	\\
2761.73	&	44125.90	&	w$^{5}$P$^{o}$	&	1	&	7927.47	&	a$^{5}$D	&	2	&	\nodata	&	43$\pm$4	&	-0.83	\\
2764.35	&	44259.36	&	w$^{5}$P$^{o}$	&	3	&	8095.21	&	a$^{5}$D	&	3	&	\nodata	&	25.9$\pm$2.1	&	-0.68	\\
2769.90	&	44186.92	&	w$^{5}$P$^{o}$	&	2	&	8095.21	&	a$^{5}$D	&	3	&	\nodata	&	68$\pm$4	&	-0.41	\\
2780.68	&	44259.36	&	w$^{5}$P$^{o}$	&	3	&	8307.57	&	a$^{5}$D	&	4	&	\nodata	&	95$\pm$5	&	-0.11	\\
2871.62	&	42908.57	&	x$^{5}$D$^{o}$	&	4	&	8095.21	&	a$^{5}$D	&	3	&	\nodata	&	6.2$\pm$0.4	&	-1.16	\\
2879.27	&	42648.26	&	x$^{5}$D$^{o}$	&	3	&	7927.47	&	a$^{5}$D	&	2	&	\nodata	&	12.6$\pm$0.7	&	-0.96	\\
2886.99	&	42438.82	&	x$^{5}$D$^{o}$	&	2	&	7810.82	&	a$^{5}$D	&	1	&	\nodata	&	18.0$\pm$1.3	&	-0.95	\\
2889.24	&	42908.57	&	x$^{5}$D$^{o}$	&	4	&	8307.57	&	a$^{5}$D	&	4	&	\nodata	&	49.1$\pm$2.5	&	-0.26	\\
2893.25	&	42648.26	&	x$^{5}$D$^{o}$	&	3	&	8095.21	&	a$^{5}$D	&	3	&	\nodata	&	33.6$\pm$1.8	&	-0.53	\\
2894.16	&	42292.96	&	x$^{5}$D$^{o}$	&	1	&	7750.78	&	a$^{5}$D	&	0	&	\nodata	&	19.6$\pm$1.4	&	-1.13	\\
2896.75	&	42438.82	&	x$^{5}$D$^{o}$	&	2	&	7927.47	&	a$^{5}$D	&	2	&	\nodata	&	22.2$\pm$1.3	&	-0.85	\\
2899.20	&	42292.96	&	x$^{5}$D$^{o}$	&	1	&	7810.82	&	a$^{5}$D	&	1	&	\nodata	&	8.2$\pm$1.1	&	-1.51	\\
2905.49	&	42218.37	&	x$^{5}$D$^{o}$	&	0	&	7810.82	&	a$^{5}$D	&	1	&	\nodata	&	72$\pm$5	&	-1.04	\\
2909.04	&	42292.96	&	x$^{5}$D$^{o}$	&	1	&	7927.47	&	a$^{5}$D	&	2	&	\nodata	&	41.8$\pm$2.4	&	-0.80	\\
2910.90	&	42438.82	&	x$^{5}$D$^{o}$	&	2	&	8095.21	&	a$^{5}$D	&	3	&	\nodata	&	27.5$\pm$1.6	&	-0.76	\\
2911.14	&	42648.26	&	x$^{5}$D$^{o}$	&	3	&	8307.57	&	a$^{5}$D	&	4	&	\nodata	&	14.7$\pm$0.9	&	-0.88	\\
2967.64	&	41782.19	&	y$^{5}$D$^{o}$	&	4	&	8095.21	&	a$^{5}$D	&	3	&	\nodata	&	31.8$\pm$1.8	&	-0.42	\\
2971.11	&	41575.10	&	y$^{5}$D$^{o}$	&	3	&	7927.47	&	a$^{5}$D	&	2	&	\nodata	&	45.9$\pm$2.4	&	-0.37	\\
2975.48	&	41409.03	&	y$^{5}$D$^{o}$	&	2	&	7810.82	&	a$^{5}$D	&	1	&	\nodata	&	53.8$\pm$2.9	&	-0.45	\\
2980.79	&	41289.17	&	y$^{5}$D$^{o}$	&	1	&	7750.78	&	a$^{5}$D	&	0	&	\nodata	&	55$\pm$3	&	-0.66	\\
2985.85	&	41409.03	&	y$^{5}$D$^{o}$	&	2	&	7927.47	&	a$^{5}$D	&	2	&	\nodata	&	45.9$\pm$2.6	&	-0.51	\\
2986.00	&	41575.10	&	y$^{5}$D$^{o}$	&	3	&	8095.21	&	a$^{5}$D	&	3	&	\nodata	&	102$\pm$5	&	-0.02	\\
2986.13	&	41289.17	&	y$^{5}$D$^{o}$	&	1	&	7810.82	&	a$^{5}$D	&	1	&	\nodata	&	14.9$\pm$2.1	&	-1.22	\\
2986.47	&	41782.19	&	y$^{5}$D$^{o}$	&	4	&	8307.57	&	a$^{5}$D	&	4	&	\nodata	&	183$\pm$9	&	0.34	\\
2988.65	&	41043.35	&	x$^{5}$P$^{o}$	&	3	&	7593.16	&	a$^{5}$S	&	2	&	\nodata	&	35.9$\pm$2.0	&	-0.47	\\
2991.89	&	41224.80	&	y$^{5}$D$^{o}$	&	0	&	7810.82	&	a$^{5}$D	&	1	&	\nodata	&	192$\pm$10	&	-0.59	\\
2995.10	&	40971.29	&	y$^{5}$F$^{o}$	&	2	&	7593.16	&	a$^{5}$S	&	2	&	\nodata	&	30.6$\pm$2.1	&	-0.69	\\
2996.58	&	41289.17	&	y$^{5}$D$^{o}$	&	1	&	7927.47	&	a$^{5}$D	&	2	&	\nodata	&	130$\pm$7	&	-0.28	\\
2998.78	&	40930.31	&	x$^{5}$P$^{o}$	&	1	&	7593.16	&	a$^{5}$S	&	2	&	\nodata	&	39.3$\pm$2.3	&	-0.80	\\
3000.88	&	41409.03	&	y$^{5}$D$^{o}$	&	2	&	8095.21	&	a$^{5}$D	&	3	&	\nodata	&	105$\pm$5	&	-0.15	\\
3005.06	&	41575.10	&	y$^{5}$D$^{o}$	&	3	&	8307.57	&	a$^{5}$D	&	4	&	\nodata	&	60$\pm$3	&	-0.25	\\
3013.03	&	40930.31	&	x$^{5}$P$^{o}$	&	1	&	7750.78	&	a$^{5}$D	&	0	&	\nodata	&	20.8$\pm$1.4	&	-1.07	\\
3014.76	&	40971.29	&	y$^{5}$F$^{o}$	&	2	&	7810.82	&	a$^{5}$D	&	1	&	\nodata	&	130$\pm$7	&	-0.05	\\
3014.91	&	41086.26	&	y$^{5}$F$^{o}$	&	3	&	7927.47	&	a$^{5}$D	&	2	&	\nodata	&	188$\pm$11	&	0.25	\\
3015.20	&	40906.46	&	y$^{5}$F$^{o}$	&	1	&	7750.78	&	a$^{5}$D	&	0	&	\nodata	&	155$\pm$9	&	-0.20	\\
3017.57	&	41224.78	&	y$^{5}$F$^{o}$	&	4	&	8095.21	&	a$^{5}$D	&	3	&	\nodata	&	242$\pm$14	&	0.47	\\
3018.49	&	40930.31	&	x$^{5}$P$^{o}$	&	1	&	7810.82	&	a$^{5}$D	&	1	&	\nodata	&	88$\pm$5	&	-0.44	\\
3018.82	&	41043.35	&	x$^{5}$P$^{o}$	&	3	&	7927.47	&	a$^{5}$D	&	2	&	\nodata	&	23.3$\pm$1.6	&	-0.65	\\
3020.67	&	40906.46	&	y$^{5}$F$^{o}$	&	1	&	7810.82	&	a$^{5}$D	&	1	&	\nodata	&	110$\pm$7	&	-0.35	\\
3021.56	&	41393.47	&	y$^{5}$F$^{o}$	&	5	&	8307.57	&	a$^{5}$D	&	4	&	\nodata	&	272$\pm$16	&	0.61	\\
3029.16	&	40930.31	&	x$^{5}$P$^{o}$	&	1	&	7927.47	&	a$^{5}$D	&	2	&	\nodata	&	25.0$\pm$1.4	&	-0.99	\\
3030.24	&	41086.26	&	y$^{5}$F$^{o}$	&	3	&	8095.21	&	a$^{5}$D	&	3	&	\nodata	&	91$\pm$6	&	-0.06	\\
3031.35	&	40906.46	&	y$^{5}$F$^{o}$	&	1	&	7927.47	&	a$^{5}$D	&	2	&	\nodata	&	22.0$\pm$1.5	&	-1.04	\\
3034.19	&	41043.35	&	x$^{5}$P$^{o}$	&	3	&	8095.21	&	a$^{5}$D	&	3	&	\nodata	&	22.7$\pm$1.3	&	-0.66	\\
3037.04	&	41224.78	&	y$^{5}$F$^{o}$	&	4	&	8307.57	&	a$^{5}$D	&	4	&	\nodata	&	38.5$\pm$2.4	&	-0.32	\\
3040.84	&	40971.29	&	y$^{5}$F$^{o}$	&	2	&	8095.21	&	a$^{5}$D	&	3	&	\nodata	&	56$\pm$3	&	-0.41	\\
3053.87	&	41043.35	&	x$^{5}$P$^{o}$	&	3	&	8307.57	&	a$^{5}$D	&	4	&	\nodata	&	73$\pm$4	&	-0.15	\\
3351.96	&	29824.75	&	y$^{5}$P$^{o}$	&	3	&	0.00	&	a$^{7}$S	&	3	&	\nodata	&	0.111$\pm$0.015	&	-2.88	\\
3379.16	&	29584.62	&	y$^{5}$P$^{o}$	&	2	&	0.00	&	a$^{7}$S	&	3	&	\nodata	&	0.111$\pm$0.013	&	-3.02	\\
3578.68	&	27935.26	&	y$^{7}$P$^{o}$	&	4	&	0.00	&	a$^{7}$S	&	3	&	\nodata	&	152$\pm$8	&	0.42	\\
3593.48	&	27820.23	&	y$^{7}$P$^{o}$	&	3	&	0.00	&	a$^{7}$S	&	3	&	\nodata	&	151$\pm$8	&	0.31	\\
3605.32	&	27728.87	&	y$^{7}$P$^{o}$	&	2	&	0.00	&	a$^{7}$S	&	3	&	\nodata	&	151$\pm$8	&	0.17	\\
3730.80	&	26796.28	&	z$^{5}$P$^{o}$	&	2	&	0.00	&	a$^{7}$S	&	3	&	\nodata	&	0.177$\pm$0.016	&	-2.73	\\
3732.02	&	26787.50	&	z$^{5}$P$^{o}$	&	3	&	0.00	&	a$^{7}$S	&	3	&	\nodata	&	0.184$\pm$0.024	&	-2.57	\\
3742.96	&	47228.80	&	x$^{5}$G$^{o}$	&	5	&	20519.60	&	a$^{5}$G	&	6	&	\nodata	&	5.1$\pm$0.3	&	-0.93	\\
3743.54	&	47228.80	&	x$^{5}$G$^{o}$	&	5	&	20523.69	&	a$^{5}$G	&	4	&	\nodata	&	7.5$\pm$0.7	&	-0.76	\\
3743.57	&	47228.80	&	x$^{5}$G$^{o}$	&	5	&	20523.94	&	a$^{5}$G	&	5	&	\nodata	&	50.8$\pm$2.6	&	0.07	\\
3743.88	&	47222.27	&	x$^{5}$G$^{o}$	&	6	&	20519.60	&	a$^{5}$G	&	6	&	\nodata	&	71$\pm$4	&	0.29	\\
3744.49	&	47222.27	&	x$^{5}$G$^{o}$	&	6	&	20523.94	&	a$^{5}$G	&	5	&	\nodata	&	4.97$\pm$0.28	&	-0.87	\\
3748.61	&	47189.87	&	x$^{5}$G$^{o}$	&	4	&	20520.92	&	a$^{5}$G	&	3	&	\nodata	&	7.8$\pm$0.5	&	-0.83	\\
3749.00	&	47189.87	&	x$^{5}$G$^{o}$	&	4	&	20523.69	&	a$^{5}$G	&	4	&	\nodata	&	41.7$\pm$2.2	&	-0.10	\\
3749.04	&	47189.87	&	x$^{5}$G$^{o}$	&	4	&	20523.94	&	a$^{5}$G	&	5	&	\nodata	&	8.8$\pm$0.8	&	-0.78	\\
3757.16	&	47125.70	&	x$^{5}$G$^{o}$	&	3	&	20517.40	&	a$^{5}$G	&	2	&	\nodata	&	6.2$\pm$0.5	&	-1.04	\\
3757.66	&	47125.70	&	x$^{5}$G$^{o}$	&	3	&	20520.92	&	a$^{5}$G	&	3	&	\nodata	&	37.7$\pm$2.0	&	-0.25	\\
3758.05	&	47125.70	&	x$^{5}$G$^{o}$	&	3	&	20523.69	&	a$^{5}$G	&	4	&	\nodata	&	10.6$\pm$0.7	&	-0.81	\\
3768.24	&	47047.47	&	x$^{5}$G$^{o}$	&	2	&	20517.40	&	a$^{5}$G	&	2	&	\nodata	&	48.1$\pm$2.5	&	-0.29	\\
3768.74	&	47047.47	&	x$^{5}$G$^{o}$	&	2	&	20520.92	&	a$^{5}$G	&	3	&	\nodata	&	11.3$\pm$0.8	&	-0.92	\\
3883.29	&	33671.55	&	z$^{5}$D$^{o}$	&	3	&	7927.47	&	a$^{5}$D	&	2	&	\nodata	&	3.46$\pm$0.18	&	-1.26	\\
3885.22	&	33542.11	&	z$^{5}$D$^{o}$	&	2	&	7810.82	&	a$^{5}$D	&	1	&	\nodata	&	3.94$\pm$0.20	&	-1.35	\\
3886.80	&	33816.06	&	z$^{5}$D$^{o}$	&	4	&	8095.21	&	a$^{5}$D	&	3	&	\nodata	&	2.11$\pm$0.11	&	-1.37	\\
3894.04	&	33423.79	&	z$^{5}$D$^{o}$	&	1	&	7750.78	&	a$^{5}$D	&	0	&	\nodata	&	3.28$\pm$0.17	&	-1.65	\\
3902.91	&	33542.11	&	z$^{5}$D$^{o}$	&	2	&	7927.47	&	a$^{5}$D	&	2	&	\nodata	&	2.66$\pm$0.13	&	-1.52	\\
3903.17	&	33423.79	&	z$^{5}$D$^{o}$	&	1	&	7810.82	&	a$^{5}$D	&	1	&	\nodata	&	0.85$\pm$0.05	&	-2.23	\\
3908.76	&	33671.55	&	z$^{5}$D$^{o}$	&	3	&	8095.21	&	a$^{5}$D	&	3	&	\nodata	&	5.55$\pm$0.28	&	-1.05	\\
3916.25	&	33338.20	&	z$^{5}$D$^{o}$	&	0	&	7810.82	&	a$^{5}$D	&	1	&	\nodata	&	7.7$\pm$0.4	&	-1.75	\\
3919.15	&	33816.06	&	z$^{5}$D$^{o}$	&	4	&	8307.57	&	a$^{5}$D	&	4	&	\nodata	&	9.4$\pm$0.5	&	-0.71	\\
3921.02	&	33423.79	&	z$^{5}$D$^{o}$	&	1	&	7927.47	&	a$^{5}$D	&	2	&	\nodata	&	5.37$\pm$0.27	&	-1.43	\\
3928.64	&	33542.11	&	z$^{5}$D$^{o}$	&	2	&	8095.21	&	a$^{5}$D	&	3	&	\nodata	&	4.20$\pm$0.21	&	-1.31	\\
3941.48	&	33671.55	&	z$^{5}$D$^{o}$	&	3	&	8307.57	&	a$^{5}$D	&	4	&	\nodata	&	2.48$\pm$0.13	&	-1.39	\\
3963.69	&	45741.49	&	y$^{5}$H$^{o}$	&	7	&	20519.60	&	a$^{5}$G	&	6	&	112	&	118$\pm$6	&	0.62	\\
3969.06	&	45707.36	&	y$^{5}$H$^{o}$	&	6	&	20519.60	&	a$^{5}$G	&	6	&	7.4	&	6.9$\pm$0.4	&	-0.67	\\
3969.74	&	45707.36	&	y$^{5}$H$^{o}$	&	6	&	20523.94	&	a$^{5}$G	&	5	&	104	&	105$\pm$5	&	0.51	\\
3976.66	&	45663.28	&	y$^{5}$H$^{o}$	&	5	&	20523.69	&	a$^{5}$G	&	4	&	98	&	93$\pm$5	&	0.39	\\
3976.70	&	45663.28	&	y$^{5}$H$^{o}$	&	5	&	20523.94	&	a$^{5}$G	&	5	&	12.8	&	12.7$\pm$1.1	&	-0.48	\\
3983.90	&	45614.88	&	y$^{5}$H$^{o}$	&	4	&	20520.92	&	a$^{5}$G	&	3	&	N  94	&	94$\pm$5	&	0.30	\\
3984.34	&	45614.88	&	y$^{5}$H$^{o}$	&	4	&	20523.69	&	a$^{5}$G	&	4	&	15.4	&	15.8$\pm$1.1	&	-0.47	\\
3991.11	&	45566.02	&	y$^{5}$H$^{o}$	&	3	&	20517.40	&	a$^{5}$G	&	2	&	95	&	100$\pm$5	&	0.22	\\
3991.67	&	45566.02	&	y$^{5}$H$^{o}$	&	3	&	20520.92	&	a$^{5}$G	&	3	&	13.4	&	13.8$\pm$0.8	&	-0.64	\\
4025.00	&	45358.63	&	z$^{3}$H$^{o}$	&	4	&	20520.92	&	a$^{5}$G	&	3	&	\nodata	&	4.0$\pm$0.4	&	-1.05	\\
4025.45	&	45358.63	&	z$^{3}$H$^{o}$	&	4	&	20523.69	&	a$^{5}$G	&	4	&	\nodata	&	0.61$\pm$0.13	&	-1.88	\\
4026.21	&	45354.18	&	z$^{3}$H$^{o}$	&	5	&	20523.94	&	a$^{5}$G	&	5	&	\nodata	&	0.45$\pm$0.06	&	-1.92	\\
4027.09	&	45348.73	&	z$^{3}$H$^{o}$	&	6	&	20523.94	&	a$^{5}$G	&	5	&	\nodata	&	3.53$\pm$0.29	&	-0.95	\\
4254.33	&	23498.84	&	z$^{7}$P$^{o}$	&	4	&	0.00	&	a$^{7}$S	&	3	&	\nodata	&	33.0$\pm$1.7	&	-0.09	\\
4261.35	&	46958.98	&	f$^{7}$D	&	5	&	23498.84	&	z$^{7}$P$^{o}$	&	4	&	\nodata	&	6.8$\pm$0.5	&	-0.69	\\
4272.90	&	46783.06	&	f$^{7}$D	&	4	&	23386.35	&	z$^{7}$P$^{o}$	&	3	&	\nodata	&	4.2$\pm$0.4	&	-0.98	\\
4274.80	&	23386.35	&	z$^{7}$P$^{o}$	&	3	&	0.00	&	a$^{7}$S	&	3	&	\nodata	&	31.7$\pm$1.6	&	-0.22	\\
4284.72	&	46637.21	&	f$^{7}$D	&	3	&	23305.01	&	z$^{7}$P$^{o}$	&	2	&	\nodata	&	2.6$\pm$0.4	&	-1.30	\\
4289.72	&	23305.01	&	z$^{7}$P$^{o}$	&	2	&	0.00	&	a$^{7}$S	&	3	&	\nodata	&	31.0$\pm$1.5	&	-0.37	\\
4293.55	&	46783.06	&	f$^{7}$D	&	4	&	23498.84	&	z$^{7}$P$^{o}$	&	4	&	\nodata	&	2.53$\pm$0.22	&	-1.20	\\
4299.71	&	46637.21	&	f$^{7}$D	&	3	&	23386.35	&	z$^{7}$P$^{o}$	&	3	&	\nodata	&	4.7$\pm$0.4	&	-1.04	\\
4305.45	&	46524.84	&	f$^{7}$D	&	2	&	23305.01	&	z$^{7}$P$^{o}$	&	2	&	\nodata	&	7.8$\pm$1.4	&	-0.97	\\
4319.64	&	46448.60	&	f$^{7}$D	&	1	&	23305.01	&	z$^{7}$P$^{o}$	&	2	&	\nodata	&	8.6$\pm$1.1	&	-1.14	\\
4320.59	&	46524.84	&	f$^{7}$D	&	2	&	23386.35	&	z$^{7}$P$^{o}$	&	3	&	\nodata	&	2.7$\pm$0.3	&	-1.42	\\
4320.61	&	46637.21	&	f$^{7}$D	&	3	&	23498.84	&	z$^{7}$P$^{o}$	&	4	&	\nodata	&	0.52$\pm$0.12	&	-1.99	\\
4337.55	&	30858.82	&	z$^{5}$F$^{o}$	&	2	&	7810.82	&	a$^{5}$D	&	1	&	5.65	&	5.75$\pm$0.29	&	-1.09	\\
4339.44	&	30965.46	&	z$^{5}$F$^{o}$	&	3	&	7927.47	&	a$^{5}$D	&	2	&	N   6.9	&	6.9$\pm$0.3	&	-0.86	\\
4339.71	&	30787.30	&	z$^{5}$F$^{o}$	&	1	&	7750.78	&	a$^{5}$D	&	0	&	4.70	&	4.66$\pm$0.23	&	-1.40	\\
4344.50	&	31106.37	&	z$^{5}$F$^{o}$	&	4	&	8095.21	&	a$^{5}$D	&	3	&	8.4	&	8.7$\pm$0.4	&	-0.65	\\
4351.05	&	30787.30	&	z$^{5}$F$^{o}$	&	1	&	7810.82	&	a$^{5}$D	&	1	&	4.67	&	4.40$\pm$0.22	&	-1.43	\\
4351.75	&	31280.35	&	z$^{5}$F$^{o}$	&	5	&	8307.57	&	a$^{5}$D	&	4	&	10.0	&	10.6$\pm$0.5	&	-0.48	\\
4356.75	&	47228.80	&	x$^{5}$G$^{o}$	&	5	&	24282.34	&	b$^{5}$D	&	4	&	\nodata	&	2.19$\pm$0.18	&	-1.16	\\
4359.62	&	30858.82	&	z$^{5}$F$^{o}$	&	2	&	7927.47	&	a$^{5}$D	&	2	&	3.98	&	3.74$\pm$0.19	&	-1.27	\\
4364.15	&	47189.87	&	x$^{5}$G$^{o}$	&	4	&	24282.34	&	b$^{5}$D	&	4	&	\nodata	&	0.39$\pm$0.07	&	-1.99	\\
4368.27	&	47189.87	&	x$^{5}$G$^{o}$	&	4	&	24303.94	&	b$^{5}$D	&	3	&	\nodata	&	1.96$\pm$0.18	&	-1.30	\\
4371.26	&	30965.46	&	z$^{5}$F$^{o}$	&	3	&	8095.21	&	a$^{5}$D	&	3	&	2.96	&	2.70$\pm$0.14	&	-1.27	\\
4373.26	&	30787.30	&	z$^{5}$F$^{o}$	&	1	&	7927.47	&	a$^{5}$D	&	2	&	0.66	&	0.58$\pm$0.04	&	-2.30	\\
4384.96	&	31106.37	&	z$^{5}$F$^{o}$	&	4	&	8307.57	&	a$^{5}$D	&	4	&	1.63	&	1.51$\pm$0.08	&	-1.41	\\
4391.74	&	30858.82	&	z$^{5}$F$^{o}$	&	2	&	8095.21	&	a$^{5}$D	&	3	&	0.389	&	0.326$\pm$0.018	&	-2.33	\\
4412.23	&	30965.46	&	z$^{5}$F$^{o}$	&	3	&	8307.57	&	a$^{5}$D	&	4	&	0.137	&	0.105$\pm$0.006	&	-2.67	\\
4496.84	&	29824.75	&	y$^{5}$P$^{o}$	&	3	&	7593.16	&	a$^{5}$S	&	2	&	\nodata	&	3.38$\pm$0.18	&	-1.14	\\
4526.44	&	42605.81	&	z$^{5}$G$^{o}$	&	6	&	20519.60	&	a$^{5}$G	&	6	&	\nodata	&	17.7$\pm$0.9	&	-0.15	\\
4527.33	&	42605.81	&	z$^{5}$G$^{o}$	&	6	&	20523.94	&	a$^{5}$G	&	5	&	\nodata	&	2.12$\pm$0.11	&	-1.07	\\
4529.84	&	42589.25	&	z$^{5}$G$^{o}$	&	5	&	20519.60	&	a$^{5}$G	&	6	&	\nodata	&	1.31$\pm$0.09	&	-1.35	\\
4530.68	&	42589.25	&	z$^{5}$G$^{o}$	&	5	&	20523.69	&	a$^{5}$G	&	4	&	\nodata	&	3.11$\pm$0.27	&	-0.98	\\
4530.73	&	42589.25	&	z$^{5}$G$^{o}$	&	5	&	20523.94	&	a$^{5}$G	&	5	&	\nodata	&	16.0$\pm$0.8	&	-0.27	\\
4535.12	&	42564.85	&	z$^{5}$G$^{o}$	&	4	&	20520.92	&	a$^{5}$G	&	3	&	\nodata	&	3.48$\pm$0.20	&	-1.02	\\
4535.69	&	42564.85	&	z$^{5}$G$^{o}$	&	4	&	20523.69	&	a$^{5}$G	&	4	&	\nodata	&	13.8$\pm$0.7	&	-0.42	\\
4535.75	&	42564.85	&	z$^{5}$G$^{o}$	&	4	&	20523.94	&	a$^{5}$G	&	5	&	\nodata	&	2.33$\pm$0.14	&	-1.19	\\
4539.76	&	42538.81	&	z$^{5}$G$^{o}$	&	3	&	20517.40	&	a$^{5}$G	&	2	&	\nodata	&	2.96$\pm$0.16	&	-1.19	\\
4540.49	&	42538.81	&	z$^{5}$G$^{o}$	&	3	&	20520.92	&	a$^{5}$G	&	3	&	\nodata	&	14.0$\pm$0.7	&	-0.52	\\
4541.06	&	42538.81	&	z$^{5}$G$^{o}$	&	3	&	20523.69	&	a$^{5}$G	&	4	&	\nodata	&	3.25$\pm$0.18	&	-1.15	\\
4544.60	&	42515.35	&	z$^{5}$G$^{o}$	&	2	&	20517.40	&	a$^{5}$G	&	2	&	\nodata	&	16.6$\pm$0.9	&	-0.59	\\
4545.33	&	42515.35	&	z$^{5}$G$^{o}$	&	2	&	20520.92	&	a$^{5}$G	&	3	&	\nodata	&	3.21$\pm$0.18	&	-1.30	\\
4545.95	&	29584.62	&	y$^{5}$P$^{o}$	&	2	&	7593.16	&	a$^{5}$S	&	2	&	\nodata	&	2.75$\pm$0.14	&	-1.37	\\
4565.50	&	29824.75	&	y$^{5}$P$^{o}$	&	3	&	7927.47	&	a$^{5}$D	&	2	&	0.306	&	0.432$\pm$0.028	&	-2.02	\\
4580.04	&	29420.90	&	y$^{5}$P$^{o}$	&	1	&	7593.16	&	a$^{5}$S	&	2	&	\nodata	&	2.34$\pm$0.12	&	-1.66	\\
4591.39	&	29584.62	&	y$^{5}$P$^{o}$	&	2	&	7810.82	&	a$^{5}$D	&	1	&	0.95	&	1.16$\pm$0.06	&	-1.74	\\
4600.74	&	29824.75	&	y$^{5}$P$^{o}$	&	3	&	8095.21	&	a$^{5}$D	&	3	&	2.10	&	2.52$\pm$0.14	&	-1.25	\\
4613.36	&	29420.90	&	y$^{5}$P$^{o}$	&	1	&	7750.78	&	a$^{5}$D	&	0	&	2.08	&	2.31$\pm$0.12	&	-1.65	\\
4616.12	&	29584.62	&	y$^{5}$P$^{o}$	&	2	&	7927.47	&	a$^{5}$D	&	2	&	3.63	&	4.02$\pm$0.20	&	-1.19	\\
4626.17	&	29420.90	&	y$^{5}$P$^{o}$	&	1	&	7810.82	&	a$^{5}$D	&	1	&	4.64	&	4.85$\pm$0.24	&	-1.33	\\
4628.47	&	46958.98	&	f$^{7}$D	&	5	&	25359.62	&	z$^{7}$F$^{o}$	&	4	&	0.80	&	1.13$\pm$0.15	&	-1.40	\\
4633.26	&	46783.06	&	f$^{7}$D	&	4	&	25206.02	&	z$^{7}$F$^{o}$	&	3	&	2.44	&	2.7$\pm$0.3	&	-1.11	\\
4639.50	&	46637.21	&	f$^{7}$D	&	3	&	25089.20	&	z$^{7}$F$^{o}$	&	2	&	5.0	&	5.5$\pm$0.4	&	-0.91	\\
4646.15	&	29824.75	&	y$^{5}$P$^{o}$	&	3	&	8307.57	&	a$^{5}$D	&	4	&	7.9	&	8.0$\pm$0.4	&	-0.74	\\
4646.79	&	46524.84	&	f$^{7}$D	&	2	&	25010.64	&	z$^{7}$F$^{o}$	&	1	&	8.7	&	8.9$\pm$0.8	&	-0.84	\\
4651.28	&	29420.90	&	y$^{5}$P$^{o}$	&	1	&	7927.47	&	a$^{5}$D	&	2	&	3.55	&	3.56$\pm$0.18	&	-1.46	\\
4652.15	&	29584.62	&	y$^{5}$P$^{o}$	&	2	&	8095.21	&	a$^{5}$D	&	3	&	N   5.68	&	5.68$\pm$0.29	&	-1.04	\\
4654.76	&	46448.60	&	f$^{7}$D	&	1	&	24971.21	&	z$^{7}$F$^{o}$	&	0	&	14.5	&	14.7$\pm$0.9	&	-0.84	\\
4663.32	&	46448.60	&	f$^{7}$D	&	1	&	25010.64	&	z$^{7}$F$^{o}$	&	1	&	28.8	&	27.9$\pm$1.7	&	-0.56	\\
4663.82	&	46524.84	&	f$^{7}$D	&	2	&	25089.20	&	z$^{7}$F$^{o}$	&	2	&	25.9	&	24.7$\pm$1.4	&	-0.39	\\
4664.79	&	46637.21	&	f$^{7}$D	&	3	&	25206.02	&	z$^{7}$F$^{o}$	&	3	&	21.5	&	21.8$\pm$1.2	&	-0.30	\\
4666.20	&	45358.63	&	z$^{3}$H$^{o}$	&	4	&	23933.90	&	a$^{3}$H	&	4	&	\nodata	&	4.44$\pm$0.24	&	-0.88	\\
4666.48	&	46783.06	&	f$^{7}$D	&	4	&	25359.62	&	z$^{7}$F$^{o}$	&	4	&	15.8	&	15.5$\pm$0.9	&	-0.34	\\
4667.17	&	45354.18	&	z$^{3}$H$^{o}$	&	5	&	23933.90	&	a$^{3}$H	&	4	&	\nodata	&	1.55$\pm$0.11	&	-1.25	\\
4669.33	&	46958.98	&	f$^{7}$D	&	5	&	25548.64	&	z$^{7}$F$^{o}$	&	5	&	8.6	&	9.4$\pm$0.6	&	-0.47	\\
4680.47	&	46448.60	&	f$^{7}$D	&	1	&	25089.20	&	z$^{7}$F$^{o}$	&	2	&	17.1	&	17.3$\pm$1.1	&	-0.77	\\
4689.38	&	46524.84	&	f$^{7}$D	&	2	&	25206.02	&	z$^{7}$F$^{o}$	&	3	&	25.4	&	24.0$\pm$1.4	&	-0.40	\\
4692.97	&	45358.63	&	z$^{3}$H$^{o}$	&	4	&	24056.11	&	a$^{3}$H	&	5	&	\nodata	&	0.24$\pm$0.04	&	-2.15	\\
4693.95	&	45354.18	&	z$^{3}$H$^{o}$	&	5	&	24056.11	&	a$^{3}$H	&	5	&	\nodata	&	4.35$\pm$0.24	&	-0.80	\\
4695.15	&	45348.73	&	z$^{3}$H$^{o}$	&	6	&	24056.11	&	a$^{3}$H	&	5	&	\nodata	&	1.78$\pm$0.10	&	-1.12	\\
4698.47	&	46637.21	&	f$^{7}$D	&	3	&	25359.62	&	z$^{7}$F$^{o}$	&	4	&	N  33.1	&	33.1$\pm$1.8	&	-0.11	\\
4708.02	&	46783.06	&	f$^{7}$D	&	4	&	25548.64	&	z$^{7}$F$^{o}$	&	5	&	41.0	&	39.0$\pm$2.1	&	0.07	\\
4718.43	&	46958.98	&	f$^{7}$D	&	5	&	25771.40	&	z$^{7}$F$^{o}$	&	6	&	49.2	&	47.3$\pm$2.5	&	0.24	\\
4727.15	&	45348.73	&	z$^{3}$H$^{o}$	&	6	&	24200.23	&	a$^{3}$H	&	6	&	\nodata	&	5.2$\pm$0.3	&	-0.65	\\
4745.27	&	42908.57	&	x$^{5}$D$^{o}$	&	4	&	21840.84	&	a$^{5}$P	&	3	&	\nodata	&	1.39$\pm$0.22	&	-1.38	\\
4789.34	&	41393.47	&	y$^{5}$F$^{o}$	&	5	&	20519.60	&	a$^{5}$G	&	6	&	\nodata	&	12.4$\pm$1.6	&	-0.33	\\
4790.34	&	41393.47	&	y$^{5}$F$^{o}$	&	5	&	20523.94	&	a$^{5}$G	&	5	&	\nodata	&	0.88$\pm$0.13	&	-1.48	\\
4814.28	&	45663.28	&	y$^{5}$H$^{o}$	&	5	&	24897.55	&	a$^{3}$G	&	4	&	\nodata	&	1.57$\pm$0.14	&	-1.22	\\
4829.31	&	41224.78	&	y$^{5}$F$^{o}$	&	4	&	20523.69	&	a$^{5}$G	&	4	&	\nodata	&	1.22$\pm$0.19	&	-1.42	\\
4829.37	&	41224.78	&	y$^{5}$F$^{o}$	&	4	&	20523.94	&	a$^{5}$G	&	5	&	\nodata	&	9.8$\pm$1.4	&	-0.51	\\
4836.87	&	45707.36	&	y$^{5}$H$^{o}$	&	6	&	25038.61	&	a$^{3}$G	&	5	&	\nodata	&	1.8$\pm$0.3	&	-1.09	\\
4847.21	&	45663.28	&	y$^{5}$H$^{o}$	&	5	&	25038.61	&	a$^{3}$G	&	5	&	\nodata	&	0.29$\pm$0.05	&	-1.95	\\
4861.19	&	41086.26	&	y$^{5}$F$^{o}$	&	3	&	20520.92	&	a$^{5}$G	&	3	&	\nodata	&	1.43$\pm$0.23	&	-1.45	\\
4861.85	&	41086.26	&	y$^{5}$F$^{o}$	&	3	&	20523.69	&	a$^{5}$G	&	4	&	\nodata	&	7.7$\pm$1.1	&	-0.72	\\
4870.80	&	45358.63	&	z$^{3}$H$^{o}$	&	4	&	24833.86	&	a$^{3}$G	&	3	&	\nodata	&	30.2$\pm$1.5	&	-0.01	\\
4880.05	&	45663.28	&	y$^{5}$H$^{o}$	&	5	&	25177.39	&	a$^{3}$F	&	4	&	\nodata	&	0.67$\pm$0.15	&	-1.58	\\
4885.96	&	45358.63	&	z$^{3}$H$^{o}$	&	4	&	24897.55	&	a$^{3}$G	&	4	&	\nodata	&	2.33$\pm$0.14	&	-1.12	\\
4887.03	&	45354.18	&	z$^{3}$H$^{o}$	&	5	&	24897.55	&	a$^{3}$G	&	4	&	\nodata	&	30.6$\pm$1.6	&	0.08	\\
4887.68	&	40971.29	&	y$^{5}$F$^{o}$	&	2	&	20517.40	&	a$^{5}$G	&	2	&	\nodata	&	0.48$\pm$0.09	&	-2.07	\\
4888.52	&	40971.29	&	y$^{5}$F$^{o}$	&	2	&	20520.92	&	a$^{5}$G	&	3	&	\nodata	&	2.4$\pm$0.4	&	-1.36	\\
4903.22	&	40906.46	&	y$^{5}$F$^{o}$	&	1	&	20517.40	&	a$^{5}$G	&	2	&	\nodata	&	7.7$\pm$1.3	&	-1.08	\\
4920.96	&	45354.18	&	z$^{3}$H$^{o}$	&	5	&	25038.61	&	a$^{3}$G	&	5	&	\nodata	&	2.98$\pm$0.16	&	-0.92	\\
4922.28	&	45348.73	&	z$^{3}$H$^{o}$	&	6	&	25038.61	&	a$^{3}$G	&	5	&	\nodata	&	50.2$\pm$2.5	&	0.38	\\
4936.34	&	45358.63	&	z$^{3}$H$^{o}$	&	4	&	25106.34	&	a$^{3}$F	&	3	&	\nodata	&	16.9$\pm$0.9	&	-0.25	\\
4953.71	&	45358.63	&	z$^{3}$H$^{o}$	&	4	&	25177.39	&	a$^{3}$F	&	4	&	\nodata	&	0.99$\pm$0.07	&	-1.48	\\
4954.81	&	45354.18	&	z$^{3}$H$^{o}$	&	5	&	25177.39	&	a$^{3}$F	&	4	&	\nodata	&	16.6$\pm$0.9	&	-0.17	\\
5013.31	&	41782.19	&	y$^{5}$D$^{o}$	&	4	&	21840.84	&	a$^{5}$P	&	3	&	\nodata	&	5.0$\pm$0.8	&	-0.77	\\
5065.92	&	41575.10	&	y$^{5}$D$^{o}$	&	3	&	21840.84	&	a$^{5}$P	&	3	&	\nodata	&	1.55$\pm$0.25	&	-1.38	\\
5067.72	&	41575.10	&	y$^{5}$D$^{o}$	&	3	&	21847.88	&	a$^{5}$P	&	2	&	\nodata	&	3.2$\pm$0.5	&	-1.07	\\
5110.75	&	41409.03	&	y$^{5}$D$^{o}$	&	2	&	21847.88	&	a$^{5}$P	&	2	&	\nodata	&	2.4$\pm$0.4	&	-1.32	\\
5113.12	&	41409.03	&	y$^{5}$D$^{o}$	&	2	&	21856.94	&	a$^{5}$P	&	1	&	\nodata	&	1.70$\pm$0.27	&	-1.48	\\
5142.26	&	41289.17	&	y$^{5}$D$^{o}$	&	1	&	21847.88	&	a$^{5}$P	&	2	&	\nodata	&	0.96$\pm$0.19	&	-1.94	\\
5144.66	&	41289.17	&	y$^{5}$D$^{o}$	&	1	&	21856.94	&	a$^{5}$P	&	1	&	\nodata	&	3.5$\pm$0.6	&	-1.37	\\
5177.42	&	46958.98	&	f$^{7}$D	&	5	&	27649.71	&	z$^{7}$D$^{o}$	&	4	&	6.6	&	6.7$\pm$0.6	&	-0.53	\\
5184.55	&	46783.06	&	f$^{7}$D	&	4	&	27500.37	&	z$^{7}$D$^{o}$	&	3	&	11.4	&	11.3$\pm$0.7	&	-0.39	\\
5192.00	&	46637.21	&	f$^{7}$D	&	3	&	27382.18	&	z$^{7}$D$^{o}$	&	2	&	N  14.0	&	14.0$\pm$0.9	&	-0.40	\\
5200.21	&	46524.84	&	f$^{7}$D	&	2	&	27300.19	&	z$^{7}$D$^{o}$	&	1	&	13.0	&	13.0$\pm$0.9	&	-0.58	\\
5204.51	&	26801.93	&	z$^{5}$P$^{o}$	&	1	&	7593.16	&	a$^{5}$S	&	2	&	\nodata	&	52.4$\pm$2.6	&	-0.19	\\
5206.04	&	26796.28	&	z$^{5}$P$^{o}$	&	2	&	7593.16	&	a$^{5}$S	&	2	&	\nodata	&	51.9$\pm$2.6	&	0.02	\\
5208.42	&	26787.50	&	z$^{5}$P$^{o}$	&	3	&	7593.16	&	a$^{5}$S	&	2	&	\nodata	&	52.1$\pm$2.6	&	0.17	\\
5220.91	&	46448.60	&	f$^{7}$D	&	1	&	27300.19	&	z$^{7}$D$^{o}$	&	1	&	10.7	&	10.6$\pm$0.7	&	-0.89	\\
5224.07	&	46637.21	&	f$^{7}$D	&	3	&	27500.37	&	z$^{7}$D$^{o}$	&	3	&	4.0	&	4.7$\pm$0.5	&	-0.87	\\
5224.97	&	46958.98	&	f$^{7}$D	&	5	&	27825.45	&	z$^{7}$D$^{o}$	&	5	&	25.6	&	26.0$\pm$1.5	&	0.07	\\
5225.02	&	46783.06	&	f$^{7}$D	&	4	&	27649.71	&	z$^{7}$D$^{o}$	&	4	&	13.0	&	13.6$\pm$0.9	&	-0.30	\\
5225.81	&	40971.29	&	y$^{5}$F$^{o}$	&	2	&	21840.84	&	a$^{5}$P	&	3	&	\nodata	&	1.54$\pm$0.24	&	-1.50	\\
5227.74	&	40971.29	&	y$^{5}$F$^{o}$	&	2	&	21847.88	&	a$^{5}$P	&	2	&	\nodata	&	0.43$\pm$0.07	&	-2.05	\\
5230.22	&	40971.29	&	y$^{5}$F$^{o}$	&	2	&	21856.94	&	a$^{5}$P	&	1	&	\nodata	&	0.92$\pm$0.15	&	-1.73	\\
5238.96	&	40930.31	&	x$^{5}$P$^{o}$	&	1	&	21847.88	&	a$^{5}$P	&	2	&	\nodata	&	4.3$\pm$0.7	&	-1.27	\\
5241.45	&	40930.31	&	x$^{5}$P$^{o}$	&	1	&	21856.94	&	a$^{5}$P	&	1	&	\nodata	&	0.97$\pm$0.22	&	-1.92	\\
5243.36	&	46448.60	&	f$^{7}$D	&	1	&	27382.18	&	z$^{7}$D$^{o}$	&	2	&	21.1	&	21.3$\pm$1.3	&	-0.58	\\
5247.57	&	26801.93	&	z$^{5}$P$^{o}$	&	1	&	7750.78	&	a$^{5}$D	&	0	&	2.03	&	2.07$\pm$0.11	&	-1.59	\\
5254.93	&	46524.84	&	f$^{7}$D	&	2	&	27500.37	&	z$^{7}$D$^{o}$	&	3	&	18.8	&	19.0$\pm$1.2	&	-0.41	\\
5255.13	&	46958.98	&	f$^{7}$D	&	5	&	27935.26	&	y$^{7}$P$^{o}$	&	4	&	16.3	&	17.6$\pm$1.2	&	-0.10	\\
5264.16	&	26801.93	&	z$^{5}$P$^{o}$	&	1	&	7810.82	&	a$^{5}$D	&	1	&	4.52	&	4.53$\pm$0.23	&	-1.25	\\
5265.16	&	46637.21	&	f$^{7}$D	&	3	&	27649.71	&	z$^{7}$D$^{o}$	&	4	&	13.9	&	15.2$\pm$0.9	&	-0.35	\\
5265.72	&	26796.28	&	z$^{5}$P$^{o}$	&	2	&	7810.82	&	a$^{5}$D	&	1	&	0.90	&	0.95$\pm$0.05	&	-1.71	\\
5272.01	&	46783.06	&	f$^{7}$D	&	4	&	27820.23	&	y$^{7}$P$^{o}$	&	3	&	10.1	&	10.2$\pm$0.8	&	-0.42	\\
5273.46	&	46783.06	&	f$^{7}$D	&	4	&	27825.45	&	z$^{7}$D$^{o}$	&	5	&	7.6	&	8.2$\pm$0.6	&	-0.51	\\
5287.20	&	46637.21	&	f$^{7}$D	&	3	&	27728.87	&	y$^{7}$P$^{o}$	&	2	&	N   4.6	&	4.6$\pm$0.5	&	-0.87	\\
5296.69	&	26801.93	&	z$^{5}$P$^{o}$	&	1	&	7927.47	&	a$^{5}$D	&	2	&	3.45	&	3.48$\pm$0.18	&	-1.36	\\
5298.28	&	26796.28	&	z$^{5}$P$^{o}$	&	2	&	7927.47	&	a$^{5}$D	&	2	&	N   3.45	&	3.45$\pm$0.18	&	-1.14	\\
5300.74	&	26787.50	&	z$^{5}$P$^{o}$	&	3	&	7927.47	&	a$^{5}$D	&	2	&	0.281	&	0.342$\pm$0.024	&	-2.00	\\
5304.18	&	46783.06	&	f$^{7}$D	&	4	&	27935.26	&	y$^{7}$P$^{o}$	&	4	&	5.9	&	5.7$\pm$0.5	&	-0.67	\\
5312.87	&	46637.21	&	f$^{7}$D	&	3	&	27820.23	&	y$^{7}$P$^{o}$	&	3	&	9.9	&	9.5$\pm$0.6	&	-0.55	\\
5318.81	&	46524.84	&	f$^{7}$D	&	2	&	27728.87	&	y$^{7}$P$^{o}$	&	2	&	10.5	&	10.0$\pm$0.7	&	-0.67	\\
5340.47	&	46448.60	&	f$^{7}$D	&	1	&	27728.87	&	y$^{7}$P$^{o}$	&	2	&	15.5	&	14.5$\pm$1.1	&	-0.73	\\
5344.79	&	46524.84	&	f$^{7}$D	&	2	&	27820.23	&	y$^{7}$P$^{o}$	&	3	&	5.2	&	4.8$\pm$0.3	&	-0.99	\\
5345.80	&	26796.28	&	z$^{5}$P$^{o}$	&	2	&	8095.21	&	a$^{5}$D	&	3	&	5.37	&	5.23$\pm$0.27	&	-0.95	\\
5348.31	&	26787.50	&	z$^{5}$P$^{o}$	&	3	&	8095.21	&	a$^{5}$D	&	3	&	1.92	&	2.05$\pm$0.11	&	-1.21	\\
5409.77	&	26787.50	&	z$^{5}$P$^{o}$	&	3	&	8307.57	&	a$^{5}$D	&	4	&	7.1	&	7.0$\pm$0.4	&	-0.67	\\
5628.62	&	45358.63	&	z$^{3}$H$^{o}$	&	4	&	27597.22	&	b$^{3}$G	&	3	&	\nodata	&	4.2$\pm$0.4	&	-0.74	\\
5664.04	&	45354.18	&	z$^{3}$H$^{o}$	&	5	&	27703.84	&	b$^{3}$G	&	4	&	\nodata	&	3.68$\pm$0.26	&	-0.71	\\
5702.32	&	45348.73	&	z$^{3}$H$^{o}$	&	6	&	27816.88	&	b$^{3}$G	&	5	&	\nodata	&	3.41$\pm$0.24	&	-0.67	\\
5712.75	&	41782.19	&	y$^{5}$D$^{o}$	&	4	&	24282.34	&	b$^{5}$D	&	4	&	\nodata	&	2.1$\pm$0.4	&	-1.03	\\
5719.81	&	41782.19	&	y$^{5}$D$^{o}$	&	4	&	24303.94	&	b$^{5}$D	&	3	&	\nodata	&	0.59$\pm$0.12	&	-1.58	\\
5781.16	&	41575.10	&	y$^{5}$D$^{o}$	&	3	&	24282.34	&	b$^{5}$D	&	4	&	\nodata	&	2.8$\pm$0.5	&	-1.00	\\
5787.04	&	41575.10	&	y$^{5}$D$^{o}$	&	3	&	24299.89	&	b$^{5}$D	&	2	&	\nodata	&	0.81$\pm$0.22	&	-1.55	\\
5788.39	&	41575.10	&	y$^{5}$D$^{o}$	&	3	&	24303.94	&	b$^{5}$D	&	3	&	\nodata	&	0.91$\pm$0.17	&	-1.49	\\
5838.65	&	41409.03	&	y$^{5}$D$^{o}$	&	2	&	24286.54	&	b$^{5}$D	&	1	&	\nodata	&	0.59$\pm$0.11	&	-1.82	\\
5844.59	&	41409.03	&	y$^{5}$D$^{o}$	&	2	&	24303.94	&	b$^{5}$D	&	3	&	\nodata	&	0.67$\pm$0.15	&	-1.77	\\
5876.54	&	41289.17	&	y$^{5}$D$^{o}$	&	1	&	24277.06	&	b$^{5}$D	&	0	&	\nodata	&	0.58$\pm$0.14	&	-2.05	\\
5884.43	&	41289.17	&	y$^{5}$D$^{o}$	&	1	&	24299.89	&	b$^{5}$D	&	2	&	\nodata	&	0.88$\pm$0.21	&	-1.86	\\
6313.22	&	47228.80	&	x$^{5}$G$^{o}$	&	5	&	31393.40	&	a$^{5}$F	&	5	&	\nodata	&	0.25$\pm$0.05	&	-1.78	\\
6322.60	&	47189.87	&	x$^{5}$G$^{o}$	&	4	&	31377.96	&	a$^{5}$F	&	4	&	\nodata	&	0.30$\pm$0.06	&	-1.80	\\
8917.13	&	42589.25	&	z$^{5}$G$^{o}$	&	5	&	31377.96	&	a$^{5}$F	&	4	&	\nodata	&	0.106$\pm$0.017	&	-1.86	\\
9290.48	&	31280.35	&	z$^{5}$F$^{o}$	&	5	&	20519.60	&	a$^{5}$G	&	6	&	\nodata	&	0.29$\pm$0.04	&	-1.38	\\
9294.23	&	31280.35	&	z$^{5}$F$^{o}$	&	5	&	20523.94	&	a$^{5}$G	&	5	&	\nodata	&	0.025$\pm$0.003	&	-2.45	\\
9446.81	&	31106.37	&	z$^{5}$F$^{o}$	&	4	&	20523.69	&	a$^{5}$G	&	4	&	\nodata	&	0.051$\pm$0.008	&	-2.21	\\
9447.03	&	31106.37	&	z$^{5}$F$^{o}$	&	4	&	20523.94	&	a$^{5}$G	&	5	&	\nodata	&	0.29$\pm$0.04	&	-1.46	\\
9571.75	&	30965.46	&	z$^{5}$F$^{o}$	&	3	&	20520.92	&	a$^{5}$G	&	3	&	\nodata	&	0.052$\pm$0.007	&	-2.30	\\
9574.29	&	30965.46	&	z$^{5}$F$^{o}$	&	3	&	20523.69	&	a$^{5}$G	&	4	&	\nodata	&	0.24$\pm$0.03	&	-1.63	\\
9667.20	&	30858.82	&	z$^{5}$F$^{o}$	&	2	&	20517.40	&	a$^{5}$G	&	2	&	\nodata	&	0.045$\pm$0.006	&	-2.51	\\
9670.49	&	30858.82	&	z$^{5}$F$^{o}$	&	2	&	20520.92	&	a$^{5}$G	&	3	&	\nodata	&	0.23$\pm$0.03	&	-1.79	\\
9734.52	&	30787.30	&	z$^{5}$F$^{o}$	&	1	&	20517.40	&	a$^{5}$G	&	2	&	\nodata	&	0.27$\pm$0.04	&	-1.95	\\
\enddata
\end{deluxetable}

\newpage
\tablenum{4}
\tablecolumns{7}
\tablewidth{0pt}

\begin{deluxetable}{lcccccccc}
\tablecaption{EW Measurements for the Survey Stars}
\tablehead{
\colhead{$\lambda$}                &
\colhead{$\chi$}                   &
\colhead{$log (gf)$}               &
\colhead{$<EW_{Sun}>$}             &
\colhead{$<EW_{HD75732}>$}         &
\colhead{$<EW_{HD140283}>$}        &      
\colhead{$<EW_{CS22892}>$}        \\
\colhead{[\AA]}                    &
\colhead{[eV]}                     &
\colhead{}                         &
\colhead{[m\AA]}                   &
\colhead{[m\AA]}                   &
\colhead{[m\AA]}                   &
\colhead{[m\AA]}                   \\
}
\startdata
Cr I : & & & & & & \\
3018.49	&	0.97	&	-0.44	&	101.6	&	\nodata	&	\nodata	&	\nodata	\\
3578.68	&	0.00	&	0.42	&	\nodata	&	\nodata	&	70.5	&	89.9	\\
3593.48	&	0.00	&	0.31	&	\nodata	&	\nodata	&	68.5	&	86.5	\\
3732.02	&	0.00	&	-2.57	&	55.5	&	\nodata	&	\nodata	&	\nodata	\\
3916.25	&	0.97	&	-1.75	&	54.0	&	\nodata	&	\nodata	&	\nodata	\\
3984.34	&	2.54	&	-0.47	&	52.5	&	\nodata	&	\nodata	&	\nodata	\\
4025.00	&	2.54	&	-1.05	&	22.8	&	\nodata	&	\nodata	&	\nodata	\\
4254.33	&	0.00	&	-0.09	&	\nodata	&	\nodata	&	65.2	&	\nodata	\\
4274.80	&	0.00	&	-0.22	&	\nodata	&	\nodata	&	61.0	&	\nodata	\\
4289.72	&	0.00	&	-0.37	&	\nodata	&	\nodata	&	55.2	&	90.5	\\
4293.55	&	2.91	&	-1.20	&	13.2	&	\nodata	&	\nodata	&	\nodata	\\
4319.64	&	2.89	&	-1.14	&	15.1	&	44.1	&	\nodata	&	\nodata	\\
4373.26	&	0.98	&	-2.30	&	35.6	&	83.9	&	\nodata	&	\nodata	\\
4496.84	&	0.94	&	-1.14	&	82.7	&	\nodata	&	3.9	&	\nodata	\\
4529.84	&	2.54	&	-1.35	&	17.1	&	\nodata	&	\nodata	&	\nodata	\\
4535.12	&	2.54	&	-1.02	&	29.5	&	68.0	&	\nodata	&	\nodata	\\
4541.06	&	2.54	&	-1.15	&	24.3	&	\nodata	&	\nodata	&	\nodata	\\
4545.95	&	0.94	&	-1.37	&	78.2	&	131.9	&	2.8	&	\nodata	\\
4591.39	&	0.97	&	-1.74	&	61.9	&	\nodata	&	\nodata	&	\nodata	\\
4600.74	&	1.00	&	-1.25	&	77.8	&	\nodata	&	2.7	&	\nodata	\\
4613.36	&	0.96	&	-1.65	&	66.2	&	\nodata	&	\nodata	&	\nodata	\\
4616.12	&	0.98	&	-1.19	&	81.0	&	136.8	&	\nodata	&	\nodata	\\
4626.17	&	0.97	&	-1.33	&	76.1	&	\nodata	&	2.6	&	\nodata	\\
4628.47	&	3.14	&	-1.40	&	6.1	&	34.7	&	\nodata	&	\nodata	\\
4633.26	&	3.13	&	-1.11	&	9.0	&	38.2	&	\nodata	&	\nodata	\\
4639.50	&	3.11	&	-0.91	&	15.1	&	\nodata	&	\nodata	&	\nodata	\\
4646.15	&	1.03	&	-0.74	&	92.0	&	\nodata	&	\nodata	&	16.9	\\
4651.28	&	0.98	&	-1.46	&	73.7	&	132.1	&	\nodata	&	\nodata	\\
4652.15	&	1.00	&	-1.04	&	92.0	&	\nodata	&	5.4	&	9.7	\\
4689.38	&	3.13	&	-0.40	&	36.0	&	\nodata	&	\nodata	&	\nodata	\\
4695.15	&	2.98	&	-1.12	&	15.3	&	\nodata	&	\nodata	&	\nodata	\\
4708.02	&	3.17	&	0.07	&	53.8	&	92.8	&	\nodata	&	\nodata	\\
4718.43	&	3.20	&	0.24	&	62.5	&	105.7	&	\nodata	&	\nodata	\\
4745.27	&	2.71	&	-1.38	&	11.7	&	\nodata	&	\nodata	&	\nodata	\\
4789.34	&	2.54	&	-0.33	&	59.2	&	99.9	&	\nodata	&	\nodata	\\
4790.34	&	2.54	&	-1.48	&	12.5	&	48.9	&	\nodata	&	\nodata	\\
4885.96	&	3.09	&	-1.12	&	12.8	&	\nodata	&	\nodata	&	\nodata	\\
4936.34	&	3.11	&	-0.25	&	42.4	&	75.4	&	\nodata	&	\nodata	\\
4953.71	&	3.12	&	-1.48	&	4.2	&	22.6	&	\nodata	&	\nodata	\\
5177.42	&	3.43	&	-0.53	&	18.1	&	\nodata	&	\nodata	&	\nodata	\\
5200.21	&	3.38	&	-0.58	&	20.8	&	56.6	&	\nodata	&	\nodata	\\
5220.91	&	3.38	&	-0.89	&	9.8	&	\nodata	&	\nodata	&	\nodata	\\
5225.81	&	2.71	&	-1.50	&	12.8	&	\nodata	&	\nodata	&	\nodata	\\
5238.96	&	2.71	&	-1.27	&	14.9	&	47.9	&	\nodata	&	\nodata	\\
5241.45	&	2.71	&	-1.92	&	3.4	&	19.5	&	\nodata	&	\nodata	\\
5243.36	&	3.39	&	-0.58	&	18.6	&	58.9	&	\nodata	&	\nodata	\\
5247.57	&	0.96	&	-1.59	&	77.6	&	125.2	&	\nodata	&	\nodata	\\
5255.13	&	3.46	&	-0.10	&	34.3	&	\nodata	&	\nodata	&	\nodata	\\
5265.16	&	3.43	&	-0.35	&	24.5	&	\nodata	&	\nodata	&	\nodata	\\
5287.20	&	3.44	&	-0.87	&	9.8	&	33.5	&	\nodata	&	\nodata	\\
5296.69	&	0.98	&	-1.36	&	87.6	&	144.9	&	\nodata	&	8.0	\\
5300.74	&	0.98	&	-2.00	&	54.2	&	103.3	&	\nodata	&	\nodata	\\
5304.18	&	3.46	&	-0.67	&	14.4	&	42.9	&	\nodata	&	\nodata	\\
5318.81	&	3.44	&	-0.67	&	13.7	&	45.5	&	\nodata	&	\nodata	\\
5340.47	&	3.44	&	-0.73	&	12.9	&	\nodata	&	\nodata	&	\nodata	\\
5345.80	&	1.00	&	-0.95	&	107.2	&	187.7	&	5.4	&	13.1	\\
5348.31	&	1.00	&	-1.21	&	93.4	&	161.3	&	3.2	&	85.0	\\
5409.77	&	1.03	&	-0.67	&	125.7	&	\nodata	&	10.0	&	20.5	\\
5628.62	&	3.42	&	-0.74	&	14.0	&	44.2	&	\nodata	&	\nodata	\\
5712.75	&	3.01	&	-1.03	&	14.7	&	51.1	&	\nodata	&	\nodata	\\
5719.81	&	3.01	&	-1.58	&	4.2	&	20.8	&	\nodata	&	\nodata	\\
5781.16	&	3.01	&	-1.00	&	14.1	&	46.7	&	\nodata	&	\nodata	\\
5844.59	&	3.01	&	-1.77	&	3.8	&	21.5	&	\nodata	&	\nodata	\\
Cr II : & & & & & & \\
3382.69	&	2.45	&	-0.98	&	100.3	&	\nodata	&	39.6	&	55.1	\\
3391.44	&	2.45	&	-1.40	&	83.9	&	\nodata	&	22.4	&	31.9	\\
3393.85	&	3.10	&	-0.99	&	82.6	&	\nodata	&	16.7	&	\nodata	\\
3408.81	&	2.48	&	-0.42	&	\nodata	&	\nodata	&	54.0	&	67.0	\\
3511.84	&	2.48	&	-1.46	&	73.3	&	\nodata	&	13.7	&	29.4	\\
3585.52	&	2.71	&	-1.39	&	83.2	&	\nodata	&	\nodata	&	\nodata	\\
3715.18	&	3.10	&	-1.37	&	\nodata	&	\nodata	&	6.5	&	11.1	\\
4558.65	&	4.07	&	-0.66	&	76.8	&	81.8	&	6.8	&	\nodata	\\
4588.20	&	4.07	&	-0.83	&	70.4	&	75.5	&	5.1	&	\nodata	\\
4592.05	&	4.07	&	-1.42	&	47.4	&	\nodata	&	0.9	&	\nodata	\\
4634.08	&	4.07	&	-0.98	&	59.4	&	\nodata	&	1.7	&	1.9	\\
4848.25	&	3.86	&	-1.00	&	60.7	&	70.4	&	2.3	&	3.7	\\
\enddata
\end{deluxetable}

\newpage
\tablenum{5}
\tablecolumns{5}
\tablewidth{0pt}

\begin{deluxetable}{lcccc}
\tablecaption{Solar Photospheric Cr I and Cr II Abundances for Different Models\tablenotemark{a}}
\tablehead{
\colhead{Model}                        & 
\colhead{$log\epsilon(Cr I)_{\sun}$}   &
\colhead{$\sigma$}                     &
\colhead{$log\epsilon(Cr II)_{\sun}$}  &
\colhead{$\sigma$}                     \\
}
\startdata
Holweger-M\"{u}ller (1974)         & $5.64\pm0.01$ & 0.07 & $5.77\pm0.03$ & 0.13\\
ATLAS (Kurcuz 1993)                & $5.52\pm0.01$ & 0.08 & $5.69\pm0.03$ & 0.13\\
Asplund et al. (2004)              & $5.49\pm0.01$ & 0.08 & $5.70\pm0.03$ & 0.13\\
Grevesse \& Sauval (1999)          & $5.58\pm0.01$ & 0.09 & $5.74\pm0.03$ & 0.13\\
NEW MARCS (Gustafsson \etal\ 2003) & $5.53\pm0.01$ & 0.08 & $5.67\pm0.03$ & 0.13\\
MARCS (Gustafsson \etal\ 1975)     & $5.52\pm0.01$ & 0.07 & $5.68\pm0.03$ & 0.13\\
\enddata

\tablenotetext{a} {Barklem damping constants and a \vt\ of 0.80 \kmsec\ are used in all of the models.}
\end{deluxetable}

\newpage
\begin{landscape}
\tablenum{6}
\tablecolumns{13}
\tablewidth{0pt}
\tabletypesize{\tiny}

\begin{deluxetable}{lcccccccccccc}
\tablecaption{Comparison to the Blackwell et al. 1987 Solar Abundances\tablenotemark{a}}
\tablehead{
\colhead{$\lambda$}                          & 
\colhead{$\chi$}                             &
\colhead{Upper}                              &
\colhead{Upper}                              &
\colhead{Lower}                              &
\colhead{Lower}                              &
\colhead{$log(gf)_{Blackwell}$}              &
\colhead{$log(gf)_{Sobeck}$}                 &
\colhead{EW$_{Moore}$}                       &
\colhead{EW$_{Blackwell}$}                   &
\colhead{EW$_{Sobeck}$}                      &
\colhead{$log\epsilon_{Blackwell}$}          &
\colhead{$log\epsilon_{Sobeck}$}            \\
\colhead{[\AA]}                              &
\colhead{[eV]}                               &
\colhead{Term}                               &
\colhead{J}                                  &
\colhead{Term}                               &
\colhead{J}                                  &
\colhead{}                                   &
\colhead{}                                   &
\colhead{[m\AA]}                             &
\colhead{[m\AA]}                             &
\colhead{[m\AA]}                             &
\colhead{}                                   &
\colhead{}                                   \\
}
\startdata
5247.57	& 0.96	& z$^{5}$P$^{o}$ & 1 & a$^{5}$D & 0 & -1.59   & -1.63	& 76      & 80.1	&	77.6	&	5.81	& 5.71\\
5264.16	& 0.97	& z$^{5}$P$^{o}$ & 1 & a$^{5}$D & 1 & \nodata & -1.25	& \nodata & \nodata	&	\nodata	&	\nodata	& \nodata\\
5265.72	& 0.97	& z$^{5}$P$^{o}$ & 2 & a$^{5}$D & 1 & \nodata & -1.71	& \nodata & \nodata	&	\nodata	&	\nodata	& \nodata\\
5296.69	& 0.98	& z$^{5}$P$^{o}$ & 1 & a$^{5}$D & 2 & -1.36   & -1.39	& \nodata & \nodata	&	87.6	&	\nodata	& 5.70\\
5298.28	& 0.98	& z$^{5}$P$^{o}$ & 2 & a$^{5}$D & 2 & -1.14   & -1.17	& \nodata & \nodata	&	\nodata	&	\nodata	& \nodata\\
5300.74	& 0.98	& z$^{5}$P$^{o}$ & 3 & a$^{5}$D & 2 & -2.00   & -2.13	& 56      & 56.3	&	54.2	&	5.78	& 5.60\\
5345.80	& 1.00	& z$^{5}$P$^{o}$ & 2 & a$^{5}$D & 3 & -0.95   & -0.98	& 107     & 116.9	&	107.2	&	5.85	& 5.66\\
5348.31	& 1.00	& z$^{5}$P$^{o}$ & 3 & a$^{5}$D & 3 & -1.21   & -1.29	& 92      & \nodata	&	93.4	&	\nodata	& 5.67\\
5409.77	& 1.03	& z$^{5}$P$^{o}$ & 3 & a$^{5}$D & 4 & -0.67   & -0.72	& \nodata & \nodata	&	125.7	&	\nodata	& 5.64\\
\enddata

\tablenotetext{a} {For abundance derivation, both studies employed the Holweger-M\"{u}ller model.  Note that
                   Blackwell \etal\ used a slightly higher \vt\ of 0.85 \kmsec.}
\end{deluxetable}
\end{landscape}

\end{document}